\documentclass[final,3p]{elsarticle}
 \usepackage{graphics}
 \usepackage{graphicx}
 \usepackage{epsfig}
\usepackage{amssymb}
 \usepackage{amsthm}
 \usepackage{lineno}
 \usepackage{amsmath}
   \numberwithin{equation}{section}
\usepackage{mathrsfs}

\newtheorem{thm}{Theorem}[section]

\newtheorem{lem}[thm]{Lemma}
\newtheorem{prop}[thm]{Proposition}
\newtheorem{defn}[thm]{Definition}

\biboptions{sort&compress,square}
\begin{document}
\begin{frontmatter}
\author[rvt1,rvt2]{Jian Wang}
\ead{wangj484@nenu.edu.cn}
\author[rvt1]{Yong Wang\corref{cor2}}
\ead{wangy581@nenu.edu.cn}
\cortext[cor2]{Corresponding author.}
\address[rvt1]{School of Mathematics and Statistics, Northeast Normal University,
Changchun, 130024, P.R.China}
\address[rvt2]{Chengde Petroleum College,
Chengde, 067000, P.R.China}

\title{Twisted Dirac Operators and the Noncommutative Residue \\ for  Manifolds with Boundary}
\begin{abstract}
In this paper, we give two Lichnerowicz  type formulas for Dirac operators and signature operators
twisted by a vector bundle with a non-unitary connection.
We also prove two Kastler-Kalau-Walze type theorems for twisted  Dirac operators and twisted  signature operators on
 4-dimensional manifolds with (resp. without) boundary.
\end{abstract}
\end{frontmatter}
\section{Introduction}
\label{1}

The noncommutative residue found in \cite{Gu,Wo} plays a prominent role in noncommutative geometry.
For one-dimensional manifolds, the noncommutative residue was discovered by Adler\cite{MA}
 in connection with geometric aspects of nonlinear partial differential equations. For arbitrary closed compact
n-dimensional manifolds, the noncommutative reside was introduced by Wodzicki in\cite{Wo} using the theory of zeta
 functions of elliptic pseudodifferential operators.
 In \cite{Co1}, Connes used the noncommutative residue to derive a conformal 4-
dimensional Polyakov action analogy. In \cite{Co2}, Connes proved
that the noncommutative residue on a compact manifold
$M$ coincided with Dixmier's trace on pseudodifferential
operators of order -dim$M$. Indeed,  Alain Connes made a challenging observation that
  the Wodzicki residue of the inverse square of the (Atiyah-Singer-Lichnerowicz) Dirac operator yields the
  Einstein-Hilbert action of general relativity, which is called the Kastler-Kalau-Walze  theorem now.
  Kastler\cite{Ka} gave a
brute-force proof of this theorem. Kalau and Walze\cite{KW} proved
this theorem in the normal coordinates system simultaneously.
Ackermann\cite{Ac} gave a note on a new proof of this theorem
by means of the heat kernel expansion.

Based on the theory of the noncommutative reside  introduced by Wodzicki, Fedosov etc.\cite{FGLS} constructed a noncommutative
residue on the algebra of classical elements in Boutet de Monvel's calculus on a compact manifold with boundary of dimension $n>2$.
For Dirac operators and signature operators on manifolds with boundary, Wang\cite{Wa3} gave an operator-theoretic explanation of the gravitational
action for manifolds with boundary and proved a Kastler-Kalau-Walze type theorem.
In \cite{Wa4},  Wang computed the lower
dimensional volume ${\rm Vol}^{(2,2)}$ for $5$-dimensional and $6$-dimensional spin manifolds with boundary and  also got a
Kastler-Kalau-Walze type theorem in this case. Then, we got a Kastler-Kalau-Walze type theorem associated with nonminimal operators
by heat equation asymptotics on compact manifolds without boundary and
proved  Kastler-Kalau-Walze type theorems for foliations with or without boundary
associated with sub-Dirac operators in \cite{WW} and \cite{WW1} respectively.

 On the other hand, Bismut and Zhang \cite{BZ} introduced the de-Rham Hodge operator twisted by a flat vector bundle with a non-metric connection,
 and extended the famous Cheeger-M$\ddot{u}$ller theorem to the non-unitary case. In  \cite{MZ}, Ma and Zhang extended
 the Atiyah-Patodi-Singer $\eta$-Invariant to the twisted non-unitary flat vector bundle case.
  In \cite{ZW}, Zhang considered the sub-signature operators twisted by a non-unitary flat vector  bundle and proved the associated
  Riemann-Roch theorem. Motivated by \cite{BZ}, \cite{MZ},  \cite{ZW} and \cite{Wa3}, we shall prove two Lichnerowicz  type formulas
  for $\widetilde{D}^{*}_{F}\widetilde{D}_{F}$ and $\hat{D}^{*}_{F}\hat{D}_{F}$(for definitions, see Section 2 and Section 4), and prove two
  Kastler-Kalau-Walze type theorems for $\widetilde{D}^{*}_{F}\widetilde{D}_{F}$ and $\hat{D}^{*}_{F}\hat{D}_{F}$ on
 manifolds with (resp. without) boundary.

This paper is organized as follows: In Section 2, we give a Lichnerowicz  type formula
  for $\widetilde{D}^{*}_{F}\widetilde{D}_{F}$. A Kastler-Kalau-Walze type theorem for $\widetilde{D}^{*}_{F}\widetilde{D}_{F}$ is given in Section 3.
   In Section 4 and Section 5, we give a Lichnerowicz  type formula and prove  a Kastler-Kalau-Walze type theorem
  for $\hat{D}^{*}_{F}\hat{D}_{F}$.

\section{ A Lichnerowicz formula  for Dirac operators twisted by a  vector bundle with a non-unitary connection}

 In this section we consider a $n$-dimensional oriented Riemannian manifold $(M, g^{M})$ equipped
with a fixed spin structure. We recall twisted Dirac operators.
Let $S(TM)$ be the spinors bundle and $F$ be an additional smooth vector bundle  equipped with a non-unitary connection $\widetilde{\nabla}^{F}$.
Let $\widetilde{\nabla}^{F,\ast}$ be the dual connection  on $F$, and define
 \begin{equation}
\nabla^{F}=\frac{\widetilde{\nabla}^{F}+\widetilde{\nabla}^{F,\ast}}{2},~~\Phi=\frac{\widetilde{\nabla}^{F}-\widetilde{\nabla}^{F,\ast}}{2},
\end{equation}
then $\nabla^{F}$ is a metric connection and $\Phi$ is an endomorphism of $F$ with a 1-form coefficient.
 We consider the tensor product vector bundle $S(TM)\otimes F$,
which becomes a Clifford module via the definition:
\begin{equation}
c(a)=c(a)\otimes \texttt{id}_{F},~~~~~a\in TM,
\end{equation}
and which we equip with the compound connection:
 \begin{equation}
\widetilde{\nabla}^{ S(TM)\otimes F}= \nabla^{ S(TM)}\otimes \texttt{id}_{ F}+ \texttt{id}_{ S(TM)}\otimes \widetilde{\nabla}^{F}.
\end{equation}
The corresponding twisted Dirac operator $\widetilde{D}_{F}$ is locally specified as follows:
 \begin{equation}
\widetilde{D}_{F}=\sum_{i=1}^{n}c(e_{i})\widetilde{\nabla}^{ S(TM)\otimes F}_{e_{i}}.
\end{equation}
Let
 \begin{equation}
\nabla^{ S(TM)\otimes F}=\nabla^{ S(TM)}\otimes \texttt{id}_{ F}+ \texttt{id}_{ S(TM)}\otimes \nabla^{F},
\end{equation}
then the spinor connection  $\widetilde{\nabla}$  induced by $\nabla^{ S(TM)\otimes F}$ is locally
given by
 \begin{equation}
\widetilde{\nabla}^{ S(TM)\otimes F}=\nabla^{ S(TM)}\otimes \texttt{id}_{ F}+ \texttt{id}_{ S(TM)}\otimes \nabla^{F}+\texttt{id}_{ S(TM)}\otimes \Phi.
\end{equation}
Let
 \begin{equation}
D_{F}=\sum_{i=1}^{n}c(e_{i})\nabla^{ S(TM)\otimes F}_{e_{i}},
\end{equation}
then the  twisted Dirac operators $\widetilde{D}_{F}$, $\widetilde{D}^{*}_{F}$ associated to the  connection
$\widetilde{\nabla}$ as follows.
 \begin{defn}
For sections $\psi\otimes \chi\in S(TM)\otimes F$,
\begin{eqnarray}
&&\widetilde{D}_{F}(\psi\otimes \chi)=D_{F}(\psi\otimes \chi)+\sum_{i=1}^{n}c(e_{i})\otimes \Phi(e_{i})(\psi\otimes \chi),\\
&&\widetilde{D}^{*}_{F}(\psi\otimes \chi)=D_{F}(\psi\otimes \chi)-\sum_{i=1}^{n}c(e_{i})\otimes \Phi^{*}(e_{i})(\psi\otimes \chi).
\end{eqnarray}
Here $\Phi^{*}(e_{i})$ denotes the adjoint of $\Phi(e_{i})$.
\end{defn}

We first establish the main theorem in this section. Let
$c(\Phi)=\sum_{i=1}^{n}c(e_{i})\otimes \Phi(e_{i})$, $c(\Phi^{*})=\sum_{i=1}^{n}c(e_{i})\otimes \Phi^{*}(e_{i})$,
 one has  the following Lichnerowicz formula.

\begin{thm}
The following identity holds:
 \begin{eqnarray}
\widetilde{D}^{*}_{F}\widetilde{D}_{F}&=&-\Big[g^{ij}(\overline{\nabla}_{\partial_{i}}\overline{\nabla}_{\partial_{j}}-
\overline{\nabla}_{\nabla^{L}_{\partial_{i}}\partial_{j}})\Big]  +\frac{1}{4}s+\frac{1}{2}\sum_{i\neq j} R^{F}(e_{i},e_{j})c(e_{i})c(e_{j})
   -c(\Phi^{*})c(\Phi)\nonumber\\
         && +\frac{1}{2}\sum_{j}\Big(\nabla_{e_{j}}^{F}c(\Phi^{*})\Big)c(e_{j})+\frac{1}{2}\sum_{j}c(e_{j})\nabla_{e_{j}}^{F}c(\Phi)
          +\frac{1}{4}\sum_{i}\Big[c(\Phi^{*})c(e_{i})-c(e_{i})c(\Phi) \Big]^{2},
\end{eqnarray}
where $s$ is the scaler curvature,  $R^{F}$ denotes the curvature-tensor of $\nabla^{F} $ on $F$, and for $X\in \Gamma(M,TM)$
 \begin{equation}
\overline{\nabla}_{X}=\nabla^{ S(TM)\otimes F}_{X}+\frac{1}{2}[c(\Phi^{*})c(X)-c(X)c(\Phi)].
\end{equation}
\end{thm}

In order to prove Theorem 2.2, we recall the basic notions of Laplace type operators \cite{PBG}.
Let $V$ be a vector bundle on $M$. Any differential operator $P$ of Laplace type has locally the form
 \begin{equation}
P=-\big(g^{ij}\partial_{i}\partial_{j}+A^{i}\partial_{i}+B\big)
\end{equation}
where $\partial_{i}$  is a natural local frame on $TM$ ,
 and $(g^{ij})_{1\leq i,j\leq n}$ is the inverse matrix associated with the metric
matrix  $(g_{ij})_{1\leq i,j\leq n}$ on $M$,
 and $A^{i}$ and $B$ are smooth sections
of $\texttt{End}(V)$ on $M$ (Endomorphism). If $P$ is a Laplace type
operator of the form (2.12), then there is a unique
connection $\nabla$ on $V$ and a unique Endomorphism $E$ such that
 \begin{equation}
P=-\Big[g^{ij}(\nabla_{\partial_{i}}\nabla_{\partial_{j}}-
 \nabla_{\nabla^{L}_{\partial_{i}}\partial_{j}})+E\Big],
\end{equation}
where $\nabla^{L}$ denotes the Levi-Civita connection on $M$. Moreover
(with local frames of $T^{*}M$ and $V$), $\nabla_{\partial_{i}}=\partial_{i}+\omega_{i} $
and $E$ are related to $g^{ij}$, $A^{i}$, and $B$ through
 \begin{eqnarray}
&&\omega_{i}=\frac{1}{2}g_{ij}\big(A^{i}\partial^{j}+g^{kl}\Gamma_{ kl}^{j} \texttt{Id}\big),\\
&&E=B-g^{ij}\big(\partial_{i}(\omega_{i})+\omega_{i}\omega_{j}-\omega_{k}\Gamma_{ ij}^{k} \big),
\end{eqnarray}
where $\Gamma_{ kl}^{j}$ is the  Christoffel coefficient of $\nabla^{L}$.

The next task then is to prove $\widetilde{D}^{*}_{F}\widetilde{D}_{F}$ has the Laplace type form.  The twisted Dirac
operator $D_{F}$ is locally given as follows.

\begin{lem}
Let  $\{e_{i}\}(1\leq i,j\leq n)$ $(\{\partial_{i}\})$ be the orthonormal
frames (natural frames respectively ) on  $TM$,
 \begin{equation}
D_{F}=\sum_{i,j}g^{ij}c(\partial_{i})\nabla^{ S(TM)\otimes F}_{\partial_{j}}=\sum_{j}^{n}c(e_{j})\nabla^{ S(TM)\otimes F}_{e_{j}},
\end{equation}
where $\nabla^{S(TM)\otimes F}_{\partial_{j}}=\partial_{j}+\sigma_{j}^{s}+\sigma_{j}^{F}$ and
$\sigma_{j}^{s}=\frac{1}{4}\sum_{j,k}\langle \nabla^{L}_{\partial_{i}}e_{j}, e_{k}\rangle c(e_{j})c(e_{k})$,
$\sigma_{j}^{F}$ is the connection  matrix of $\nabla^{F}$.
\end{lem}

Let $\partial^{j}=g^{ij}\partial_{i}, \sigma^{i}=g^{ij}\sigma_{j}, \Gamma^{k}=g^{ij}\Gamma_{ij}^{k}$.
From (6a) in \cite{Ka}, we have
\begin{eqnarray}
D_{F}^{2}&=&-g^{ij}\partial_{i}\partial_{j}-2\sigma^{j}_{S(TM)\otimes F}\partial_{j}+\Gamma^{k}\partial_{k}
                        -g^{ij}\Big[\partial_{i}(\sigma^{j}_{S(TM)\otimes F}) +\sigma^{i}_{S(TM)\otimes F}\sigma^{j}_{S(TM)\otimes F}
                  \nonumber\\
                  && -\Gamma_{ij}^{k}\sigma_{S(TM)\otimes F}^{k}\Big]+\frac{1}{4}s+\frac{1}{2}\sum_{i\neq j} R^{F}(e_{i},e_{j})c(e_{i})c(e_{j}).
\end{eqnarray}
where $s$ is the scaler curvature.

We will computer $\widetilde{D}^{*}_{F}\widetilde{D}_{F}$. We note that
 \begin{equation}
\widetilde{D}^{*}_{F}\widetilde{D}_{F}=D_{F}^{2}-c(\Phi^{*})D_{F}+D_{F}c(\Phi)-c(\Phi^{*})c(\Phi),
\end{equation}
and
\begin{eqnarray}
-c(\Phi^{*})D_{F}+D_{F}c(\Phi)&=&-\sum_{j}c(\Phi^{*})c(e_{j})\Big[e_{j}+ \sigma^{S(TM)\otimes F}_{j}\Big]
+\sum_{j}c(e_{j})\otimes c(\Phi)e_{j}+\sum_{j}c(e_{j})\otimes e_{j}\big(c(\Phi)\big)\nonumber\\
 &&
+\sum_{j}\Big[c(e_{j})\sigma_{j}^{S(TM)}\otimes c(\Phi)+c(e_{j})\otimes \sigma_{j}^{ F}c(\Phi) \Big].
\end{eqnarray}
Combining (2.17)-(2.19), we obtain the specification of $\widetilde{D}^{*}_{F}\widetilde{D}_{F}$.
\begin{eqnarray}
\widetilde{D}^{*}_{F}\widetilde{D}_{F}&=&-g^{ij}\partial_{i}\partial_{j}-2\sigma^{j}_{S(TM)\otimes F}\partial_{j}+\Gamma^{k}\partial_{k}
                -\sum_{j}\Big[c(\Phi^{*})c(e_{j})-c(e_{j})\otimes c(\Phi) \Big]e_{j}\nonumber\\
                &&-g^{ij}\Big[\partial_{i}(\sigma^{j}_{S(TM)\otimes F}) +\sigma^{i}_{S(TM)\otimes F}\sigma^{j}_{S(TM)\otimes F}
                  -\Gamma_{ij}^{k}\sigma_{S(TM)\otimes F}^{k}\Big]\nonumber\\
                &&-\sum_{j}\Big[c(\Phi^{*})c(e_{j}) \Big]\sigma^{S(TM)\otimes F}_{j}
               +\sum_{j}c(e_{j})\otimes e_{j}\big(c(\Phi)\big)\nonumber\\
                &&+\sum_{j}\Big[c(e_{j})\sigma_{j}^{S(TM)}\otimes c(\Phi)+c(e_{j})\otimes \sigma_{j}^{ F}c(\Phi) \Big]-c(\Phi^{*})c(\Phi)\nonumber\\
                               &&+\frac{1}{4}s+\frac{1}{2}\sum_{i\neq j} R^{F}(e_{i},e_{j})c(e_{i})c(e_{j}).
\end{eqnarray}
In terms of local coordinates $\{\partial_{i}\}$ inducing the  coordinate transformation
 $e_{j}=\sum_{k=1}^{n}\langle e_{j}, \texttt{d}x^{k}\rangle \partial_{k}$, let $\Gamma^{k}=g^{ij}\Gamma_{ij}^{k} $,  then
 \begin{equation}
\omega_{j}=\sigma^{j}_{S(TM)\otimes F}+\frac{1}{2}\Big(\sum_{k=1}^{n}\langle e_{k}, \texttt{d}x^{j}\rangle c(\Phi^{*})c(e_{k})
                 -\sum_{k=1}^{n}\langle e_{k}, \texttt{d}x^{j}\rangle c(e_{k})c(\Phi)\Big)+\frac{1}{2}\Gamma^{i}.
\end{equation}
For a smooth vector field $X\in \Gamma(M,TM)$, let $c(X)$ denote the Clifford action. By direct computation in normal coordinates, we obtain
 \begin{equation}
\overline{\nabla}_{X}=\nabla^{ S(TM)\otimes F}_{X}+\frac{1}{2}[c(\Phi^{*})c(X)-c(X)c(\Phi)].
\end{equation}

 We now compute $E$.
Regrouping the terms and inserting (2.21) into (2.15), we obtain
\begin{eqnarray}
E&=&g^{ij}\Big[\partial_{i}(\sigma^{j}_{S(TM)\otimes F}) +\sigma^{i}_{S(TM)\otimes F}\sigma^{j}_{S(TM)\otimes F}
                  -\Gamma_{ij}^{k}\sigma_{S(TM)\otimes F}^{k}\Big]
                  +\sum_{j}\Big[c(\Phi^{*})c(e_{j}) \Big]\sigma^{S(TM)\otimes F}_{j}\nonumber\\
                &&-\sum_{j}c(e_{j})\otimes e_{j}\big(c(\Phi)\big)
                -\sum_{j}\Big[c(e_{j})\sigma_{j}^{S(TM)}\otimes c(\Phi)+c(e_{j})\otimes \sigma_{j}^{ F}c(\Phi) \Big]\nonumber\\
                &&+c(\Phi^{*})c(\Phi)-\frac{1}{4}s-\frac{1}{2}\sum_{i\neq j} R^{F}(e_{i},e_{j})c(e_{i})c(e_{j})\nonumber\\
                 &&-\partial^{j}\big(\sigma^{j}_{S(TM)\otimes F}\big)-\frac{1}{2}\partial^{j}\Big(\sum_{k=1}^{n}\langle e_{k}, \texttt{d}x^{j}\rangle
                  c(\Phi^{*})c(e_{k})-\sum_{k=1}^{n}\langle e_{k}, \texttt{d}x^{j}\rangle c(e_{k})c(\Phi)\Big)\nonumber\\
                  &&-g^{ij}\sigma^{i}_{S(TM)\otimes F}\sigma^{j}_{S(TM)\otimes F}
                  -\frac{1}{2}g^{ij}\sigma^{i}_{S(TM)\otimes F}\Big(\sum_{k=1}^{n}\langle e_{k}, \texttt{d}x^{j}\rangle
                  c(\Phi^{*})c(e_{k})-\sum_{k=1}^{n}\langle e_{k}, \texttt{d}x^{j}\rangle c(e_{k})c(\Phi)\Big)\nonumber\\
                   &&-\frac{1}{2}g^{ij}\Big(\sum_{k=1}^{n}\langle e_{k}, \texttt{d}x^{i}\rangle
                  c(\Phi^{*})c(e_{k})-\sum_{k=1}^{n}\langle e_{k}, \texttt{d}x^{i}\rangle c(e_{k})c(\Phi)\Big)\sigma^{j}_{S(TM)\otimes F}\nonumber\\
                    &&-\frac{1}{4}g^{ij}\Big(\sum_{k=1}^{n}\langle e_{k}, \texttt{d}x^{i}\rangle
     c(\Phi^{*})c(e_{k})-\sum_{k=1}^{n}\langle e_{k}, \texttt{d}x^{i}\rangle c(e_{k})c(\Phi)\Big)\nonumber\\
  &&\times\Big(\sum_{k=1}^{n}\langle e_{k}, \texttt{d}x^{j}\rangle
                  c(\Phi^{*})c(e_{k})-\sum_{k=1}^{n}\langle e_{k}, \texttt{d}x^{j}\rangle c(e_{k})c(\Phi)\Big)\nonumber\\
                  &&+\Big[\sigma^{k}_{S(TM)\otimes F}+\frac{1}{2}\Big(\sum_{l=1}^{n}\langle e_{l}, \texttt{d}x^{k}\rangle c(\Phi^{*})c(e_{l})
                 -\sum_{l=1}^{n}\langle e_{l}, \texttt{d}x^{k}\rangle c(e_{l})c(\Phi)\Big)\Big]\Gamma^{k}.
\end{eqnarray}

Since $E$ is globally defined on $M$, so we can perform
computations of $E$ in normal coordinates. In terms of normal coordinates about $x_{0}$ one has:
$\sigma^{j}_{S(TM)}(x_{0})=0$ $e_{j}\big(c(e_{i})\big)(x_{0})=0$, $\Gamma^{k}(x_{0})=0$, we conclude that
\begin{eqnarray}
E(x_{0})&=&-\frac{1}{4}s-\frac{1}{2}\sum_{i\neq j} R^{F}(e_{i},e_{j})c(e_{i})c(e_{j})
           -\frac{1}{4}\sum_{i}\Big[c(\Phi^{*})c(e_{i})-c(e_{i})c(\Phi) \Big]^{2}+c(\Phi^{*})c(\Phi)\nonumber\\
            && -\frac{1}{2}\sum_{j}\Big[e_{j}\big(c(\Phi^{*})\big)c(e_{j}) +c(e_{j}) e_{j}\big(c(\Phi)\big)  \Big]
               -\frac{1}{2}\sum_{j}\Big[\sigma_{j}^{ F}, c(\Phi^{*})\Big]c(e_{j})-\frac{1}{2}\sum_{j}c(e_{j})\Big[\sigma_{j}^{ F}, c(\Phi)\Big]\nonumber\\
       &=&-\frac{1}{4}s-\frac{1}{2}\sum_{i\neq j} R^{F}(e_{i},e_{j})c(e_{i})c(e_{j})
         -\frac{1}{4}\sum_{i}\Big[c(\Phi^{*})c(e_{i})-c(e_{i})c(\Phi) \Big]^{2}+c(\Phi^{*})c(\Phi) \nonumber\\
        && -\frac{1}{2}\sum_{j}\Big(\nabla_{e_{j}}^{F}c(\Phi^{*})\Big)c(e_{j})-\frac{1}{2}\sum_{j}c(e_{j})\nabla_{e_{j}}^{F}c(\Phi).
\end{eqnarray}
which, together with (2.13), yields Theorem 2.2.

From  Theorem 1 in \cite{Ka} and Theorem 1 in \cite{KW},
for $M$ a compact $n$ dimensional ($n\geq 4$, even) Riemannian manifold and $\widetilde{D}^{*}_{F}\widetilde{D}_{F}$
a generalized Laplacian acting on sections of  vector bundle  on $M$, the
following relation holds:
\begin{equation}
Wres(\widetilde{D}^{*}_{F}\widetilde{D}_{F})^{(\frac{-n+2}{2})}=\frac{(2\pi)^{\frac{n}{2}}}{(\frac{n}{2}-2)!}\int_{M}{\bf{Tr}}(\frac{s}{6}+E)\texttt{d}vol_{M}
\end{equation}
where Wres denotes the noncommutative residue.

By (2.24) and $Tr(e_{i}e_{j}) =0~(i\neq j)$,  we find for the trace
\begin{eqnarray}
\texttt{Tr}(E)&=&\texttt{Tr}\Big[-\frac{1}{4}s
         +c(\Phi^{*})c(\Phi)-\frac{1}{4}\sum_{i}\big[c(\Phi^{*})c(e_{i})-c(e_{i})c(\Phi) \big]^{2}\nonumber\\
         &&~~~~-\frac{1}{2}\sum_{j}\nabla_{e_{j}}^{F}\big(c(\Phi^{*})\big)c(e_{j})-\frac{1}{2}\sum_{j}c(e_{j})\nabla_{e_{j}}^{F}
            \big(c(\Phi)\big)\Big].
\end{eqnarray}
Substituting (2.26) into (2.25), we obtain
\begin{thm}
For even $n$-dimensional compact spin manifolds
without boundary, the following equality
holds:
\begin{eqnarray}
Wres(\widetilde{D}^{*}_{F}\widetilde{D}_{F})^{(\frac{-n+2}{2})}&=&\frac{(2\pi)^{\frac{n}{2}}}{(\frac{n}{2}-2)!}\int_{M}{\bf{Tr}}
 \Big[-\frac{s}{12}+c(\Phi^{*})c(\Phi)-\frac{1}{4}\sum_{i}\big[c(\Phi^{*})c(e_{i})-c(e_{i})c(\Phi) \big]^{2}\nonumber\\
         &&~~~~-\frac{1}{2}\sum_{j}\nabla_{e_{j}}^{F}\big(c(\Phi^{*})\big)c(e_{j})-\frac{1}{2}\sum_{j}c(e_{j})\nabla_{e_{j}}^{F}
            \big(c(\Phi)\big)\Big]\texttt{d}vol_{M}.
\end{eqnarray}
where $s$ is the scaler curvature.
\end{thm}

 \section{A Kastler-Kalau-Walze type theorem for $4$-dimensional manifolds with boundary  associated with twisted Dirac Operators }
In this section, we shall prove a Kastler-Kalau-Walze type formula for 4-dimensional compact manifolds with boundary.
Some basic facts and formulae about Boutet de Monvel's calculus are recalled as follows.

Let $$ F:L^2({\bf R}_t)\rightarrow L^2({\bf R}_v);~F(u)(v)=\int e^{-ivt}u(t)\texttt{d}t$$ denote the Fourier transformation and
$\varphi(\overline{{\bf R}^+}) =r^+\varphi({\bf R})$ (similarly define $\varphi(\overline{{\bf R}^-}$)), where $\varphi({\bf R})$
denotes the Schwartz space and
  \begin{equation}
r^{+}:C^\infty ({\bf R})\rightarrow C^\infty (\overline{{\bf R}^+});~ f\rightarrow f|\overline{{\bf R}^+};~
 \overline{{\bf R}^+}=\{x\geq0;x\in {\bf R}\}.
\end{equation}
We define $H^+=F(\varphi(\overline{{\bf R}^+}));~ H^-_0=F(\varphi(\overline{{\bf R}^-}))$ which are orthogonal to each other. We have the following
 property: $h\in H^+~(H^-_0)$ iff $h\in C^\infty({\bf R})$ which has an analytic extension to the lower (upper) complex
half-plane $\{{\rm Im}\xi<0\}~(\{{\rm Im}\xi>0\})$ such that for all nonnegative integer $l$,
 \begin{equation}
\frac{\texttt{d}^{l}h}{\texttt{d}\xi^l}(\xi)\sim\sum^{\infty}_{k=1}\frac{\texttt{d}^l}{\texttt{d}\xi^l}(\frac{c_k}{\xi^k})
\end{equation}
as $|\xi|\rightarrow +\infty,{\rm Im}\xi\leq0~({\rm Im}\xi\geq0)$.

 Let $H'$ be the space of all polynomials and $H^-=H^-_0\bigoplus H';~H=H^+\bigoplus H^-.$ Denote by $\pi^+~(\pi^-)$ respectively the
 projection on $H^+~(H^-)$. For calculations, we take $H=\widetilde H=\{$rational functions having no poles on the real axis$\}$ ($\tilde{H}$
 is a dense set in the topology of $H$). Then on $\tilde{H}$,
 \begin{equation}
\pi^+h(\xi_0)=\frac{1}{2\pi i}\lim_{u\rightarrow 0^{-}}\int_{\Gamma^+}\frac{h(\xi)}{\xi_0+iu-\xi}\texttt{d}\xi,
\end{equation}
where $\Gamma^+$ is a Jordan close curve included ${\rm Im}\xi>0$ surrounding all the singularities of $h$ in the upper half-plane and
$\xi_0\in {\bf R}$. Similarly, define $\pi^{'}$ on $\tilde{H}$,
 \begin{equation}
\pi'h=\frac{1}{2\pi}\int_{\Gamma^+}h(\xi)\texttt{d}\xi.
\end{equation}
So, $\pi'(H^-)=0$. For $h\in H\bigcap L^1(R)$, $\pi'h=\frac{1}{2\pi}\int_{R}h(v)\texttt{d}v$ and for $h\in H^+\bigcap L^1(R)$, $\pi'h=0$.
Denote by $\mathcal{B}$ Boutet de Monvel's algebra (for details, see \cite{Wa1}).

In the following, we will compute  $\widetilde{Wres}[\pi^{+}(\widetilde{D}_{F}^{*})^{-1} \circ\pi^{+}\widetilde{D}_{F}^{-1}]$.

Let $M$ be a compact manifold with boundary $\partial M$. We assume that the metric $g^{M}$ on $M$ has
the following form near the boundary
 \begin{equation}
 g^{M}=\frac{1}{h(x_{n})}g^{\partial M}+\texttt{d}x _{n}^{2} ,
\end{equation}
where $g^{\partial M}$ is the metric on $\partial M$. Let $U\subset
M$ be a collar neighborhood of $\partial M$ which is diffeomorphic $\partial M\times [0,1)$. By the definition of $h(x_n)\in C^{\infty}([0,1))$
and $h(x_n)>0$, there exists $\tilde{h}\in C^{\infty}((-\varepsilon,1))$ such that $\tilde{h}|_{[0,1)}=h$ and $\tilde{h}>0$ for some
sufficiently small $\varepsilon>0$. Then there exists a metric $\hat{g}$ on $\hat{M}=M\bigcup_{\partial M}\partial M\times
(-\varepsilon,0]$ which has the form on $U\bigcup_{\partial M}\partial M\times (-\varepsilon,0 ]$
 \begin{equation}
\hat{g}=\frac{1}{\tilde{h}(x_{n})}g^{\partial M}+\texttt{d}x _{n}^{2} ,
\end{equation}
such that $\hat{g}|_{M}=g$.
We fix a metric $\hat{g}$ on the $\hat{M}$ such that $\hat{g}|_{M}=g$.
Note $\widetilde{D}_{F}$ is the  twisted Dirac operator on the spinor bundle $S(TM)\otimes F$ corresponding to the
connection $\widetilde{\nabla}$.

Now we recall the main theorem in \cite{FGLS}.

\begin{thm}\label{th:32}{\bf(Fedosov-Golse-Leichtnam-Schrohe)}
 Let $X$ and $\partial X$ be connected, ${\rm dim}X=n\geq3$,
 $A=\left(\begin{array}{lcr}\pi^+P+G &   K \\
T &  S    \end{array}\right)$ $\in \mathcal{B}$ , and denote by $p$, $b$ and $s$ the local symbols of $P,G$ and $S$ respectively.
 Define:
 \begin{eqnarray}
{\rm{\widetilde{Wres}}}(A)&=&\int_X\int_{\bf S}{\rm{tr}}_E\left[p_{-n}(x,\xi)\right]\sigma(\xi)dx \nonumber\\
&&+2\pi\int_ {\partial X}\int_{\bf S'}\left\{{\rm tr}_E\left[({\rm{tr}}b_{-n})(x',\xi')\right]+{\rm{tr}}
_F\left[s_{1-n}(x',\xi')\right]\right\}\sigma(\xi')dx',
\end{eqnarray}
Then~~ a) ${\rm \widetilde{Wres}}([A,B])=0 $, for any
$A,B\in\mathcal{B}$;~~ b) It is a unique continuous trace on
$\mathcal{B}/\mathcal{B}^{-\infty}$.
\end{thm}

  Denote by $\sigma_{l}(A)$ the $l$-order symbol of an operator A. An application of (2.1.4) in \cite{Wa1} shows that
\begin{equation}
\widetilde{Wres}[\pi^{+}(\widetilde{D}^{*}_{F})^{-1} \circ\pi^{+}\widetilde{D}_{F}^{-1}]=\int_{M}
\int_{|\xi|=1}{\bf{trace}}_{S(TM)\otimes F}
  [\sigma_{-n}\big( (\widetilde{D}^{*}_{F})^{-1} \circ \widetilde{D}_{F}^{-1}\big)]\sigma(\xi)\texttt{d}x+\int_{\partial M}\Psi,
\end{equation}
where
 \begin{eqnarray}
\Psi&=&\int_{|\xi'|=1}\int_{-\infty}^{+\infty}\sum_{j,k=0}^{\infty}\sum \frac{(-i)^{|\alpha|+j+k+\ell}}{\alpha!(j+k+1)!}
{\bf{trace}}_{S(TM)\otimes F}\Big[\partial_{x_{n}}^{j}\partial_{\xi'}^{\alpha}\partial_{\xi_{n}}^{k}\sigma_{r}^{+}
(( \widetilde{D}^{*}_{F})^{-1})(x',0,\xi',\xi_{n})\nonumber\\
&&\times\partial_{x_{n}}^{\alpha}\partial_{\xi_{n}}^{j+1}\partial_{x_{n}}^{k}\sigma_{l}(\widetilde{D}_{F}^{-1})(x',0,\xi',\xi_{n})\Big]
\texttt{d}\xi_{n}\sigma(\xi')\texttt{d}x' ,
\end{eqnarray}
and the sum is taken over $r-k+|\alpha|+\ell-j-1=-n,r\leq-1,\ell\leq-1$.

Locally we can use Theorem 2.4 to compute the interior term of (3.8), then
 \begin{eqnarray}
&&\int_M\int_{|\xi|=1}{\rm trace}_{S(TM)\otimes F}[\sigma_{-4}((\widetilde{D}^{*}_{F})^{-1} \circ \widetilde{D}_{F}^{-1})]\sigma(\xi)dx\nonumber\\
 &=& 4 \pi^{2}\int_{M}{\bf{Tr}}
 \Big[-\frac{s}{12}+c(\Phi^{*})c(\Phi)
-\frac{1}{4}\sum_{i}\big[c(\Phi^{*})c(e_{i})-c(e_{i})c(\Phi) \big]^{2}
     -\frac{1}{2}\sum_{j}\nabla_{e_{j}}^{F}\big(c(\Phi^{*})\big)c(e_{j})\nonumber\\
     &&-\frac{1}{2}\sum_{j}c(e_{j})\nabla_{e_{j}}^{F}
            \big(c(\Phi)\big)\Big]\texttt{d}vol_{M}.
\end{eqnarray}

So we only need to compute $\int_{\partial M}\Psi$.
Let us now turn to compute the symbol expansion of $\widetilde{D}_{F}^{-1}$. Recall the definition of the twisted  Dirac operator
 $\widetilde{D}_{F}$ in Definition 2.1.
Let $\nabla^{TM}$
denote the Levi-civita connection about $g^M$. In the local coordinates $\{x_i; 1\leq i\leq n\}$ and the fixed orthonormal frame
$\{\widetilde{e_1},\cdots,\widetilde{e_n}\}$, the connection matrix $(\omega_{s,t})$ is defined by
\begin{equation}
\nabla^{TM}(\widetilde{e_1},\cdots,\widetilde{e_n})= (\widetilde{e_1},\cdots,\widetilde{e_n})(\omega_{s,t}).
\end{equation}
Let $c(\widetilde{e_i})$ denote the Clifford action. Let $g^{ij}=g(dx_i,dx_j)$ and
\begin{equation}
\nabla^{TM}_{\partial_i}\partial_j=\sum_k\Gamma_{ij}^k\partial_k; ~\Gamma^k=g^{ij}\Gamma_{ij}^k.
\end{equation}
Let the cotangent vector $\xi=\sum \xi_jdx_j$ and $\xi^j=g^{ij}\xi_i$.
 By Lemma 1 in \cite{Wa4} and Lemma 2.1 in \cite{Wa3}, for any fixed point $x_0\in\partial M$, we can choose the normal coordinates $U$
 of $x_0$ in $\partial M$ (not in $M$). By the composition formula and (2.2.11) in \cite{Wa3}, we obtain
\begin{lem}
 Let $\widetilde{D}^{*}_{F},  \widetilde{D}_{F}$ be the twisted  Dirac operators on  $\Gamma(S(TM)\otimes F)$.
Then
\begin{eqnarray}
&&\sigma_{-1}(( \widetilde{D}^{*}_{F} )^{-1})=\sigma_{-1}(\widetilde{D}_{F}^{-1})=\frac{\sqrt{-1}c(\xi)}{|\xi|^2}; \\
&&\sigma_{-2}((\widetilde{D}^{*}_{F})^{-1})=\frac{c(\xi)\sigma_0(\widetilde{D}^{*}_{F})c(\xi)}{|\xi|^4}+\frac{c(\xi)}{|\xi|^6}\sum_jc(dx_j)
\Big[\partial_{x_j}[c(\xi)]|\xi|^2-c(\xi)\partial_{x_j}(|\xi|^2)\Big] ;\\
&& \sigma_{-2}(\widetilde{D}_{F}^{-1})=\frac{c(\xi)\sigma_0(\widetilde{D}_{F})c(\xi)}{|\xi|^4}+\frac{c(\xi)}{|\xi|^6}\sum_jc(dx_j)
\Big[\partial_{x_j}[c(\xi)]|\xi|^2-c(\xi)\partial_{x_j}(|\xi|^2)\Big],
\end{eqnarray}
where
\begin{eqnarray}
\sigma_0(\widetilde{D}^{*}_{F})&=& \sigma_{0}(D)+\sum_{j=1}^{n}c(e_{j})\big(\sigma_{j}^{F}-\Phi^{*}(e_{j})\big);\\
\sigma_0(\widetilde{D}_{F})&=& \sigma_{0}(D)+\sum_{j=1}^{n}c(e_{j})\big(\sigma_{j}^{F}+\Phi(e_{j})\big).
\end{eqnarray}
\end{lem}

Let us now turn to compute $\Psi$ (see formula (3.9) for definition of $\Psi$). Since the sum is taken over $-r-\ell+k+j+|\alpha|=3,
 \ r, \ell\leq-1$, then we have the boundary term of (3.8) is the sum of  the following five terms.

 {\bf case a)~I)}~$r=-1,~l=-1~k=j=0,~|\alpha|=1$

From (3.9) we have
 \begin{equation}
{\rm case~a)~I)}=-\int_{|\xi'|=1}\int^{+\infty}_{-\infty}\sum_{|\alpha|=1}
{\rm trace} \Big[\partial^\alpha_{\xi'}\pi^+_{\xi_n}\sigma_{-1}((\widetilde{D}^{*}_{F})^{-1})\times
\partial^\alpha_{x'}\partial_{\xi_n}\sigma_{-1}(\widetilde{D}_{F}^{-1})\Big](x_0)\texttt{d}\xi_n\sigma(\xi')\texttt{d}x'.
\end{equation}
By Lemma 2.2 in \cite{Wa3}, for $i<n$, then
 \begin{equation}
\partial_{x_i}\sigma_{-1}(\widetilde{D}_{F}^{-1})(x_0)=\partial_{x_i}\left(\frac{\sqrt{-1}c(\xi)}{|\xi|^2}\right)(x_0)=
\frac{\sqrt{-1}\partial_{x_i}[c(\xi)](x_0)}{|\xi|^2}
-\frac{\sqrt{-1}c(\xi)\partial_{x_i}(|\xi|^2)(x_0)}{|\xi|^4}=0,
\end{equation}
so \textbf{case a) I)} vanishes.

{\bf case a)~II)}~$r=-1,~l=-1~k=|\alpha|=0,~j=1$

From (3.9) we have
 \begin{equation}
{\rm case \
a)~II)}=-\frac{1}{2}\int_{|\xi'|=1}\int^{+\infty}_{-\infty} {\rm
trace} \Big[\partial_{x_n}\pi^+_{\xi_n}\sigma_{-1}((\widetilde{D}^{*}_{F})^{-1})\times
\partial_{\xi_n}^2\sigma_{-1}(\widetilde{D}_{F}^{-1})\Big](x_0)\texttt{d}\xi_n\sigma(\xi')\texttt{d}x',
\end{equation}
An application of Lemma 2.1 and Lemma 2.2 in \cite{Wa3} shows that
  \begin{equation}
\partial^2_{\xi_n}\sigma_{-1}(\widetilde{D}_{F}^{-1})=\sqrt{-1}\left(-\frac{6\xi_nc(\texttt{d}x_n)+2c(\xi')}
{|\xi|^4}+\frac{8\xi_n^2c(\xi)}{|\xi|^6}\right);
\end{equation}
and
  \begin{equation}
\partial_{x_n}\sigma_{-1}((\widetilde{D}^{*}_{F})^{-1})(x_0)
=\frac{\sqrt{-1}\partial_{x_n}c(\xi')(x_0)}{|\xi|^2}-\frac{\sqrt{-1}c(\xi)|\xi'|^2h'(0)}{|\xi|^4}.
\end{equation}
By (2.1) in \cite{Wa3} and the Cauchy integral formula, we obtain
\begin{eqnarray}
\pi^+_{\xi_n}\Big[\frac{c(\xi)}{|\xi|^4}\Big](x_0)|_{|\xi'|=1}
&=&\pi^+_{\xi_n}\Big[\frac{c(\xi')+\xi_nc(\texttt{d}x_n)}{(1+\xi_n^2)^2}\Big]\nonumber\\
&=&\frac{1}{2\pi i}\lim_{u\rightarrow 0^-}\int_{\Gamma^+}\frac{\frac{c(\xi')+\eta_nc(\texttt{d}x_n)}{(\eta_n+i)^2(\xi_n+iu-\eta_n)}}
{(\eta_n-i)^2}\texttt{d}\eta_n \nonumber\\
&=&\Big[\frac{c(\xi')+\eta_nc(\texttt{d}x_n)}{(\eta_n+i)^2(\xi_n-\eta_n)}\Big]^{(1)}|_{\eta_n=i}\nonumber\\
&=&-\frac{ic(\xi')}{4(\xi_n-i)}-\frac{c(\xi')+ic(\texttt{d}x_n)}{4(\xi_n-i)^2}.
\end{eqnarray}
Similarly,
 \begin{equation}
\pi^+_{\xi_n}\left[\frac{\sqrt{-1}\partial_{x_n}c(\xi')}{|\xi|^2}\right](x_0)|_{|\xi'|=1}
=\frac{\partial_{x_n}[c(\xi')](x_0)}{2(\xi_n-i)}.
\end{equation}
Combining (3.23) and (3.24), we have
 \begin{equation}
\pi^+_{\xi_n}\partial_{x_n}\sigma_{-1}(D_{F}^{-1})(x_0)|_{|\xi'|=1}=\frac{\partial_{x_n}[c(\xi')](x_0)}{2(\xi_n-i)}+\sqrt{-1}h'(0)
\left[\frac{ic(\xi')}{4(\xi_n-i)}+\frac{c(\xi')+ic(\texttt{d}x_n)}{4(\xi_n-i)^2}\right].
\end{equation}
By the relation of the Clifford action and ${\rm tr}{AB}={\rm tr }{BA}$, we have the equalities:
\begin{eqnarray}
&&{\rm tr}[c(\xi')c(\texttt{d}x_n)]=0;~~{\rm tr}[c(\texttt{d}x_n)^2]=-4 \texttt{dim}F;~~{\rm tr}[c(\xi')^2](x_0)|_{|\xi'|=1}=-4 \texttt{dim}F;\nonumber\\
&&{\rm tr}[\partial_{x_n}c(\xi')c(\texttt{d}x_n)]=0;~~{\rm tr}[\partial_{x_n}c(\xi')c(\xi')](x_0)|_{|\xi'|=1}=-2h'(0) \texttt{dim}F.
\end{eqnarray}
From equation (3.21), (3.25) and (3.26), one sees that
\begin{eqnarray}
{\rm case~ a)~II)}&=&-\int_{|\xi'|=1}\int^{+\infty}_{-\infty}\frac{i h'(0)(\xi_n-i)^2}
{(\xi_n-i)^4(\xi_n+i)^3}\texttt{d}\xi_n\sigma(\xi')\texttt{d}x'\nonumber\\
&=&-ih'(0)\Omega_3\int_{\Gamma^+}\frac{1}{(\xi_n-i)^2(\xi_n+i)^3}\texttt{d}\xi_n\texttt{d}x'\nonumber\\
&=&-ih'(0)\Omega_32\pi i[\frac{1}{(\xi_n+i)^3}]^{(1)}|_{\xi_n=i}\texttt{d}x'\nonumber\\
&=&-\frac{3}{8}\pi h'(0) \texttt{dim}F\Omega_3\texttt{d}x'.
\end{eqnarray}

 {\bf case a)~III)}~$r=-1,~l=-1~j=|\alpha|=0,~k=1$

From (3.9) we have
 \begin{equation}
{\rm case~ a)~III)}=-\frac{1}{2}\int_{|\xi'|=1}\int^{+\infty}_{-\infty}
{\rm trace} \Big[\partial_{\xi_n}\pi^+_{\xi_n}\sigma_{-1}((\widetilde{D}^{*}_{F})^{-1})\times
\partial_{\xi_n}\partial_{x_n}\sigma_{-1}(\widetilde{D}_{F}^{-1})\Big](x_0)\texttt{d}\xi_n\sigma(\xi')\texttt{d}x',
\end{equation}
Then an application of Lemma 2.2 in \cite{Wa3} shows
  \begin{equation}
\partial_{\xi_n}\partial_{x_n}q_{-1}(x_0)|_{|\xi'|=1}=-\sqrt{-1}h'(0)
\left[\frac{c(\texttt{d}x_n)}{|\xi|^4}-4\xi_n\frac{c(\xi')+\xi_nc(\texttt{d}x_n)}{|\xi|^6}\right]-
\frac{2\xi_n\sqrt{-1}\partial_{x_n}c(\xi')(x_0)}{|\xi|^4},
\end{equation}
and
  \begin{equation}
\partial_{\xi_n}\pi^+_{\xi_n}q_{-1}(x_0)|_{|\xi'|=1}=-\frac{c(\xi')+ic(\texttt{d}x_n)}{2(\xi_n-i)^2}.
\end{equation}

Similar to $\textbf{case a) II)}$, we obtain
 \begin{equation}
\textbf{case a) III)}=\frac{3}{8}\pi h'(0) \texttt{dim}F\Omega_3\texttt{d}x'.
\end{equation}

 {\bf case b)}~$r=-2,~l=-1,~k=j=|\alpha|=0$

From (3.9) we have
\begin{eqnarray}
{\rm case~ b)}&=&-i\int_{|\xi'|=1}\int^{+\infty}_{-\infty}
{\rm trace} \Big[\pi^+_{\xi_n}\sigma_{-2}((\widetilde{D}^{*}_{F})^{-1})\times
\partial_{\xi_n}\sigma_{-1}(\widetilde{D}_{F}^{-1})\Big](x_0)\texttt{d}\xi_n\sigma(\xi')\texttt{d}x' \nonumber\\
&=&-i\int_{|\xi'|=1}\int^{+\infty}_{-\infty}
{\rm trace} \Big[\pi^+_{\xi_n}\sigma_{-2}(D^{-1})\times
\partial_{\xi_n}\sigma_{-1}(\widetilde{D}_{F}^{-1})\Big](x_0)\texttt{d}\xi_n\sigma(\xi')\texttt{d}x'\nonumber\\
&&-i\int_{|\xi'|=1}\int_{-\infty}^{+\infty}\text{trace}\Big[\pi^+_{\xi_n}\Big( \frac{c(\xi)\alpha c(\xi)}{|\xi|^4}\Big)
\times \partial_{\xi_{n}}\sigma_{-1}(\widetilde{D}_{F}^{-1})\Big](x_{0})\texttt{d}\xi_{n}\sigma(\xi')\texttt{d}x'.
\end{eqnarray}
where  $\alpha=\sum_{j=1}^{4}c(e_{j})(\sigma_{j}^{F}-\Phi^{*}(e_{j}))$.

From (3.45) in \cite{WW1}, we have
 \begin{equation}
-i\int_{|\xi'|=1}\int^{+\infty}_{-\infty}
{\rm trace} [\pi^+_{\xi_n}\sigma_{-2}(D^{-1})\times
\partial_{\xi_n}\sigma_{-1}(\widetilde{D}_{F}^{-1})](x_0)\texttt{d}\xi_n\sigma(\xi')\texttt{d}x'=\frac{9}{8} h'(0)\pi \texttt{dim}F\Omega_3\texttt{d}x'.
\end{equation}
On the other hand,
 \begin{equation}
\pi^+_{\xi_n}\Big[ \frac{c(\xi)\alpha c(\xi)}{|\xi|^4}\Big]=
 \frac{(-i\xi_n-2)c(\xi')\alpha c(\xi')-i\Big[c(\texttt{d}x_{n})\alpha c(\xi')+c(\xi')\alpha c(\texttt{d}x_{n})\Big]
  -i\xi_n c(\texttt{d}x_{n})\alpha c(\texttt{d}x_{n})  }{4(\xi_n-i)^{2}}.
\end{equation}
By the relation of the Clifford action and $\texttt{tr}AB=\texttt{tr}BA$, then we have the equalities
\begin{eqnarray}
&&\texttt{tr}\Big[c(\texttt{d}x_{n})\sum_{j=1}^{4}c(e_{j})(\sigma_{j}^{F}-\Phi^{*}(e_{j}))\Big]=\texttt{tr}\Big[-\texttt{id}\otimes
(\sigma_{n}^{F}-\Phi^{*}(e_{n}))\Big],   \\
&& \texttt{tr}\Big[c(\xi')\sum_{j=1}^{4}c(e_{j})(\sigma_{j}^{F}-\Phi^{*}(e_{j}))\Big]=\texttt{tr}\Big[-\sum_{j=1}^{3}\xi_j
(\sigma_{j}^{F}-\Phi^{*}(e_{j}))\Big].
\end{eqnarray}
We note that $i<n,~\int_{|\xi'|=1}\xi_i\sigma(\xi')=0$,
so (3.36) has no contribution for computing case b). Hence
\begin{eqnarray}
&&-i\int_{|\xi'|=1}\int_{-\infty}^{+\infty}\text{trace}[\pi^+_{\xi_n}\Big[ \frac{c(\xi)\alpha c(\xi)}{|\xi|^4}\Big]
\times \partial_{\xi_{n}}\sigma_{-1}(\widetilde{D}_{F}^{-1})](x_{0})\texttt{d}\xi_{n}\sigma(\xi')\texttt{d}x'\nonumber\\
&=&-\frac{1}{4}\pi\Omega_{3}\texttt{Tr}[\texttt{id}\otimes (\sigma_{n}^{F}-\Phi^{*}(e_{j}))]\texttt{d}x'.
\end{eqnarray}
Combining (3.34) and (3.37), we have
\begin{equation}
{\rm case~ b)}=\Big[\frac{9}{8} h'(0) \texttt{dim}F
-\frac{1}{4}\texttt{Tr}\Big(\texttt{id}\otimes \big(\sigma_{n}^{F}-\Phi^{*}(e_{j})\big)\Big)\Big]\pi \Omega_{3}\texttt{d}x'.
\end{equation}

{\bf  case c)}~$r=-1,~l=-2,~k=j=|\alpha|=0$

From (3.9) we have
\begin{eqnarray}
{\rm case~ c)}&=&-i\int_{|\xi'|=1}\int^{+\infty}_{-\infty}{\rm trace}\Big [\pi^+_{\xi_n}\sigma_{-1}((\widetilde{D}^{*}_{F})^{-1})\times
\partial_{\xi_n}\sigma_{-2}(\widetilde{D}_{F}^{-1})\Big](x_0)d\xi_n\sigma(\xi')dx'\nonumber\\
&=&-i\int_{|\xi'|=1}\int^{+\infty}_{-\infty}{\rm trace} \Big[\pi^+_{\xi_n}\sigma_{-1}((\widetilde{D}^{*}_{F})^{-1})\times
\partial_{\xi_n}\sigma_{-2}(D^{-1})\Big](x_0)d\xi_n\sigma(\xi')dx'\nonumber\\
&&-i\int_{|\xi'|=1}\int_{-\infty}^{+\infty}\text{trace}\Big[\pi_{\xi_{n}}^{+}\sigma_{-1}((\widetilde{D}^{*}_{F})^{-1})\times
\partial_{\xi_n}\Big( \frac{c(\xi)\beta c(\xi)}{|\xi|^4}\Big)\Big](x_{0})\texttt{d}\xi_{n}\sigma(\xi')\texttt{d}x',
\end{eqnarray}
where $\beta=\sum_{j=1}^{n}c(e_{j})(\sigma_{j}^{F}+\Phi(e_{j}))$.

From (3.50) in \cite{WW1}, we have
 \begin{equation}
-i\int_{|\xi'|=1}\int^{+\infty}_{-\infty}{\rm trace} \Big[\pi^+_{\xi_n}\sigma_{-1}((\widetilde{D}^{*}_{F})^{-1})\times
\partial_{\xi_n}\sigma_{-2}(\widetilde{D}_{F}^{-1})\Big](x_0)d\xi_n\sigma(\xi')dx'=-\frac{9}{8} h'(0)\pi \texttt{dim}F\Omega_3dx'.
\end{equation}
On the other hand,
\begin{equation}
\pi_{\xi_{n}}^{+}\sigma_{-1}((\widetilde{D}^{*}_{F})^{-1})=\frac{c(\xi')+ic(\texttt{d}x_{n})}{2(\xi_{n}-i)}.
\end{equation}
By the relation of the Clifford action and $\texttt{tr}AB=\texttt{tr}BA$, then we have the equalities
\begin{eqnarray}
&&\texttt{tr}\Big[c(\texttt{d}x_{n})\sum_{j=1}^{4}c(e_{j})(\sigma_{j}^{F}+\Phi(e_{j}))\Big]=\texttt{tr}\Big[-\texttt{id}\otimes
(\sigma_{n}^{F}+\Phi(e_{n}))\Big],   \\
&& \texttt{tr}\Big[c(\xi')\sum_{j=1}^{4}c(e_{j})(\sigma_{j}^{F}+\Phi(e_{j}))\Big]=\texttt{tr}\Big[-\sum_{j=1}^{3}\xi_j
(\sigma_{j}^{F}+\Phi(e_{j}))\Big].
\end{eqnarray}
From (3.41) and (3.42) we obtain
\begin{eqnarray}
&&-i\int_{|\xi'|=1}\int_{-\infty}^{+\infty}\text{trace}\Big[\pi_{\xi_{n}}^{+}\sigma_{-1}((\widetilde{D}^{*}_{F})^{-1})\times
\partial_{\xi_n}\Big( \frac{c(\xi)\beta c(\xi)}{|\xi|^4}\Big)\Big](x_{0})\texttt{d}\xi_{n}\sigma(\xi')\texttt{d}x' \nonumber\\
&=&\frac{1}{4}\pi\Omega_{3}\texttt{Tr}[\texttt{id}\otimes (\sigma_{n}^{F}+\Phi(e_{n}))]\texttt{d}x'.
\end{eqnarray}
Combining (3.40) and (3.44), we have
\begin{equation}
{\rm case~ c)}=\Big[-\frac{9}{8} h'(0) \texttt{dim}F+\frac{1}{4}\texttt{Tr}[\texttt{id}\otimes (\sigma_{n}^{F}+\Phi(e_{n}))]\Big]
\pi \Omega_{3}\texttt{d}x'.
\end{equation}
We note that $\texttt{dim} S(TM)=4$, now $\Psi$  is the sum of the \textbf{case (a, b, c)}, so
\begin{equation}
\sum \textbf{case a, b , c}=\texttt{Tr}_{F}\big(\Phi^{*}(e_{n})+\Phi(e_{n})\big)\pi \Omega_{3}\texttt{d}x'.
\end{equation}

Hence we conclude that
\begin{thm}
 Let M be a 4-dimensional compact manifolds   with the boundary $\partial M$. Then
 \begin{eqnarray}
\widetilde{Wres}[\pi^{+}(\widetilde{D}_{F}^{*})^{-1} \circ\pi^{+}\widetilde{D}_{F}^{-1}]
 &=& 4 \pi^{2}\int_{M}{\bf{Tr}}
 \Big[-\frac{s}{12}+c(\Phi^{*})c(\Phi)-\frac{1}{4}\sum_{i}\big[c(\Phi^{*})c(e_{i})-c(e_{i})c(\Phi) \big]^{2}\nonumber\\
         &&-\frac{1}{2}\sum_{j}\nabla_{e_{j}}^{F}\big(c(\Phi^{*})\big)c(e_{j})-\frac{1}{2}\sum_{j}c(e_{j})\nabla_{e_{j}}^{F}
            \big(c(\Phi)\big)\Big]\texttt{d}vol_{M}  \nonumber\\
 &&  +\int_{\partial_{M}}{\bf{Tr}}_{F}\big(\Phi^{*}(e_{n})+\Phi(e_{n})\big)\pi \Omega_{3}\texttt{d}x',
\end{eqnarray}
where  $s$ is the scalar curvature.
\end{thm}

\section{A Lichnerowicz formula for twisted signature operators}

Let us recall the definition of twisted signature operators. We consider a $n$-dimensional oriented Riemannian manifold $(M, g^{M})$.
Let $F$ be a real vector bundle over $M$. let $g^{F}$ be an Euclidean metric on $F$. Let
 \begin{equation}
\wedge^{\ast}(T^{\ast}M)=\bigoplus_{i=0}^{n}\wedge^{i}(T^{\ast}M)
\end{equation}
be the real exterior algebra bundle of $T^{\ast}M$. Let
 \begin{equation}
\Omega^{\ast}(M,F)=\bigoplus_{i=0}^{n}\Omega^{i}(M,F)=\bigoplus_{i=0}^{n}C^{\infty}(M,\wedge^{\ast}(T^{\ast}M)\otimes F)
\end{equation}
be the set of smooth sections of $\wedge^{\ast}(T^{\ast}M)\otimes F$. Let $\ast$ be the Hodge star operator of $g^{TM}$.
It extends  on  $\wedge^{\ast}(T^{\ast}M)\otimes F$ by acting on $F$ as identity. Then $\Omega^{\ast}(M,F)$ inherits the following
standardly induced inner product
 \begin{equation}
\langle \alpha, \beta  \rangle=\int_{M}\langle \alpha\wedge^{\ast}\beta  \rangle_{F},~~~~\alpha, \beta \in\Omega^{\ast}(M,F).
\end{equation}
Let $\widehat{\nabla}^{F}$ be the non-Euclidean connection on $F$. Let $d^{F}$ be the obvious extension of $\nabla^{F}$ on $\Omega^{\ast}(M,F)$.
Let $\delta^{F}=d^{F\ast}$ be the formal adjoint operator of $d^{F}$ with respect to the inner product. Let $\hat{D}^{F}$ be the differential
operator acting on $\Omega^{\ast}(M,F)$ defined by
 \begin{equation}
\hat{D}^{F}=d^{F}+\delta^{F}.
\end{equation}
Let
 \begin{equation}
\omega(F,g^{F})=\widehat{\nabla}^{F,\ast}-\widehat{\nabla}^{F},~~\nabla^{F,e}=\nabla^{F}+\frac{1}{2}\omega(F,g^{F}).
\end{equation}
Then $\nabla^{F,e}$ is an Euclidean connection on $(F,g^{F})$.

Let $\nabla^{\wedge^{\ast}(T^{\ast}M)}$ be the Euclidean connection on $\wedge^{\ast}(T^{\ast}M)$ induced canonically by the Levi-Civita
connection $\nabla^{TM}$ of $g^{TM}$. Let $\nabla^{e}$ be the Euclidean connection on $\wedge^{\ast}(T^{\ast}M)\otimes F$ obtained from
 the tensor product of $\nabla^{\wedge^{\ast}(T^{\ast}M)}$ and $\nabla^{F,e}$.
Let $\{e_{1},\cdots,e_{n}\}$ be an oriented (local) orthonormal basis of $TM$. The following result was proved by Proposition in \cite{BZ}.

 \begin{prop} \cite{BZ}
The following identity holds
 \begin{equation}
d^{F}+\delta^{F}=\sum_{i=1}^{n}c(e_{i})\nabla^{e}_{e_{i}}-\frac{1}{2}\sum_{i=1}^{n}\hat{c}(e_{i})\omega(F,g^{F})(e_{i}).
\end{equation}
\end{prop}
Let
 \begin{equation}
D_{F}^{e}=\sum_{j=1}^{n}c(e_{j})\nabla^{e}_{e_{j}},
\end{equation}
then the  twisted signature operators $\hat{D}_{F}$, $\hat{D}^{*}_{F}$ as follows.
 \begin{defn}\label{1}
For sections $\psi\otimes \chi\in \wedge^{\ast}(T^{\ast}M)\otimes F$,
\begin{eqnarray}\label{eq:1}
&&\hat{D}_{F}(\psi\otimes \chi)=D_{F}^{e}(\psi\otimes \chi)
 -\frac{1}{2}\sum_{i=1}^{n}\hat{c}(e_{i})\omega(F,g^{F})(e_{i})(\psi\otimes \chi),\\
 \label{eq:2}
&&\hat{D}^{*}_{F}(\psi\otimes \chi)=D_{F}^{*,e}(\psi\otimes \chi)
-\frac{1}{2}\sum_{i=1}^{n}\hat{c}(e_{i})\omega^{*}(F,g^{F})(e_{i})(\psi\otimes \chi).
\end{eqnarray}
Here $\omega^{*}(F,g^{F})(e_{i})$ denotes the adjoint of $\omega(F,g^{F})(e_{i})$.
\end{defn}

We first establish the main theorem in this section. Let  $\hat{c}(\omega)=\sum_{i}c(e_{i}) \omega(F,g^{F})(e_{i})$
 and $\hat{c}(\omega^{*})=\sum_{i}c(e_{i}) \omega^{*}(F,g^{F})(e_{i})$, then
\begin{thm}\label{th:1}
The following identity holds
 \begin{eqnarray}\label{eq:1}
\hat{D}^{*}_{F}\hat{D}_{F}&=&-\Big[g^{ij}(\nabla'_{\partial_{i}}\nabla'_{\partial_{j}}-
\nabla'_{\nabla^{L}_{\partial_{i}}\partial_{j}})\Big]  +\frac{1}{4}s+\frac{1}{2}\sum_{i\neq j} R^{F,e}(e_{i},e_{j})c(e_{i})c(e_{j})
   +\frac{1}{4}\hat{c}(\omega^{*})\hat{c}(\omega)\nonumber\\
         && +\frac{1}{4}\sum_{i}\nabla_{e_{i}}^{F}\hat{c}(\omega^{*})c(e_{i})-\frac{1}{4}\sum_{i}c(e_{i})\nabla_{e_{i}}^{F}\hat{c}(\omega)
          +\frac{1}{16}\sum_{i}\Big[\hat{c}(\omega^{*})c(e_{i})+c(e_{i})\hat{c}(\omega) \Big]^{2},
\end{eqnarray}
where $s$ is the scaler curvature,  $R^{F,e}$ denotes the curvature-tensor on $F$.
\end{thm}
Now we shall prove Theorem 4.3. Similar to (2.20), we have
\begin{eqnarray}\label{eq:3}
\hat{D}^{*}_{F}\hat{D}_{F}&=&-g^{ij}\partial_{i}\partial_{j}-2\sigma^{j}_{\wedge^{\ast}(T^{\ast}M)\otimes F}\partial_{j}+\Gamma^{k}\partial_{k}
                -\frac{1}{2}\sum_{j}\Big[\hat{c}(\omega^{*})c(e_{j})+c(e_{j})\otimes \hat{c}(\omega) \Big]e_{j}\nonumber\\
                &&-g^{ij}\Big[\partial_{i}(\sigma^{j,e}_{\wedge^{\ast}(T^{\ast}M)\otimes F}) +\sigma^{i}_{\wedge^{\ast}(T^{\ast}M)\otimes F}
                \sigma^{j,e}_{\wedge^{\ast}(T^{\ast}M)\otimes F,e}
                  -\Gamma_{ij}^{k}\sigma_{\wedge^{\ast}(T^{\ast}M)\otimes F}^{k}\Big]\nonumber\\
                &&-\frac{1}{2}\sum_{j}\hat{c}(\omega^{*})c(e_{j})\sigma^{\wedge^{\ast}(T^{\ast}M)\otimes F,e}_{j}
               -\frac{1}{2}\sum_{j}c(e_{j})\otimes e_{j}\big(\hat{c}(\omega)\big) \nonumber\\
                &&-\frac{1}{2}\sum_{j}c(e_{j})\sigma_{j}^{\wedge^{\ast}(T^{\ast}M)\otimes F,e}\otimes \hat{c}(\omega)
                +\frac{1}{4}\hat{c}(\omega^{*})\hat{c}(\omega)\nonumber\\
                               &&+\frac{1}{4}s+\frac{1}{2}\sum_{i\neq j} R^{F,e}(e_{i},e_{j})c(e_{i})c(e_{j}).
\end{eqnarray}
In terms of local coordinates $\{\partial_{i}\}$ inducing the  coordinate transformation
 $e_{j}=\sum_{k=1}^{n}\langle e_{j}, \texttt{d}x^{k}\rangle \partial_{k}$, then
 \begin{equation}
\omega_{j}=\sigma^{j}_{\wedge^{\ast}(T^{\ast}M)}+\sigma^{j,e}_{F}
+\frac{1}{4}\Big(\sum_{k=1}^{n}\langle e_{k}, \texttt{d}x^{j}\rangle \hat{c}(\omega^{*}(F,g^{F}))c(e_{k})
                 -\sum_{k=1}^{n}\langle e_{k}, \texttt{d}x^{j}\rangle c(e_{k})\hat{c}(\omega(F,g^{F}))\Big)+\frac{1}{2}\Gamma^{j}.
\end{equation}
 For a smooth vector field $X\in \Gamma(M,TM)$,  then
 \begin{equation}
\nabla'_{X}=\nabla^{ \wedge^{\ast}(T^{\ast}M)\otimes F,e}_{X}+\frac{1}{4}[\hat{c}(\omega^{*}(F,g^{F}))c(X)-c(X)\hat{c}(\omega(F,g^{F}))].
\end{equation}

Since $E$ is globally defined on $M$, so we can perform
computations of $E$ in normal coordinates. In terms of normal coordinates about $x_{0}$ one has:
$\sigma^{j}_{\wedge^{\ast}(T^{\ast}M)}(x_{0})=0$, $e_{j}\big(c(e_{i})\big)(x_{0})=0$, $\Gamma^{k}(x_{0})=0$.
From (2.15) and (4.12), we obtain
\begin{eqnarray}
E(x_{0})&=&-\frac{1}{4}s-\frac{1}{2}\sum_{i\neq j} R^{F,e}(e_{i},e_{j})c(e_{i})c(e_{j})
           -\frac{1}{16}\sum_{i}\Big[\hat{c}(\omega^{*})c(e_{i})-c(e_{i})\hat{c}(\omega) \Big]^{2}
           -\frac{1}{4}\hat{c}(\omega^{*})\hat{c}(\omega)\nonumber\\
            && -\frac{1}{4}\sum_{j}\Big[e_{j}\big(\hat{c}(\omega^{*})\big)c(e_{j})-c(e_{j}) e_{j}\big(\hat{c}(\omega)\big)  \Big]
               -\frac{1}{4}\sum_{j}\Big[\sigma_{j}^{ F}, \hat{c}(\omega^{*})\Big]c(e_{j})
               +\frac{1}{4}\sum_{j}c(e_{j})\Big[\sigma_{j}^{ F}, \hat{c}(\omega)\Big]\nonumber\\
       &=&-\frac{1}{4}s-\frac{1}{2}\sum_{i\neq j} R^{F,e}(e_{i},e_{j})c(e_{i})c(e_{j})
         -\frac{1}{16}\sum_{i}\Big[\hat{c}(\omega^{*})c(e_{i})-c(e_{i})\hat{c}(\omega) \Big]^{2}
         -\frac{1}{4}\hat{c}(\omega^{*})\hat{c}(\omega)\nonumber\\
        && -\frac{1}{4}\sum_{j}\nabla_{e_{j}}^{F}\big(\hat{c}(\omega^{*})\big)c(e_{j})
        +\frac{1}{4}\sum_{j}c(e_{j})\nabla_{e_{j}}^{F}\big(\hat{c}(\omega)\big).
\end{eqnarray}
which, together with (2.13), yields Theorem 4.3.

From (4.14) we have
\begin{eqnarray}
\texttt{Tr}(E)&=&\texttt{Tr}\Big\{
     -\frac{1}{4}s-\frac{1}{2}\sum_{i\neq j} R^{F,e}(e_{i},e_{j})c(e_{i})c(e_{j})
         -\frac{1}{16}\sum_{i}\Big[\hat{c}(\omega^{*})c(e_{i})-c(e_{i})\hat{c}(\omega) \Big]^{2}
         \nonumber\\
        && -\frac{1}{4}\hat{c}(\omega^{*})\hat{c}(\omega)-\frac{1}{4}\sum_{j}\nabla_{e_{j}}^{F}\big(\hat{c}(\omega^{*})\big)c(e_{j})
        +\frac{1}{4}\sum_{j}c(e_{j})\nabla_{e_{j}}^{F}\big(\hat{c}(\omega)\big)
                        \Big\}\nonumber\\
        &=&\texttt{Tr}\Big[-\frac{1}{4}s +\frac{n}{16}[\hat{c}(\omega^{*})-\hat{c}(\omega)]^{2}
         -\frac{1}{4}\hat{c}(\omega^{*})\hat{c}(\omega) -\frac{1}{4}\sum_{j}\nabla_{e_{j}}^{F}\big(\hat{c}(\omega^{*})\big)c(e_{j})\nonumber\\
       && +\frac{1}{4}\sum_{j}c(e_{j})\nabla_{e_{j}}^{F}\big(\hat{c}(\omega)\big) \Big].
\end{eqnarray}
Hence we conclude that
\begin{thm}
For even $n$-dimensional oriented  compact Riemainnian manifolds without boundary,
the following equality holds:
\begin{eqnarray}
Wres(\hat{D}^{*}_{F}\hat{D}_{F})^{(\frac{-n+2}{2})}&=&\frac{(2\pi)^{\frac{n}{2}}}{(\frac{n}{2}-2)!}\int_{M}{\bf{Tr}}
 \Big[-\frac{s}{12} +\frac{n}{16}[\hat{c}(\omega^{*})-\hat{c}(\omega)]^{2}
         -\frac{1}{4}\hat{c}(\omega^{*})\hat{c}(\omega)\nonumber\\
       && -\frac{1}{4}\sum_{j}\nabla_{e_{j}}^{F}\big(\hat{c}(\omega^{*})\big)c(e_{j})
        +\frac{1}{4}\sum_{j}c(e_{j})\nabla_{e_{j}}^{F}\big(\hat{c}(\omega)\big) \Big]\texttt{d}vol_{M}.
\end{eqnarray}
\end{thm}

 \section{A Kastler-Kalau-Walze theorem for $4$-dimensional Riemannian manifolds with boundary  associated to twisted Signature Operators }
In this section, we shall prove a Kastler-Kalau-Walze type formula for $\hat{D}^{*}_{F}\hat{D}_{F}$.
An application of (2.1.4) in \cite{Wa1} shows that
\begin{equation}
\widetilde{Wres}[\pi^{+}(\hat{D}^{*}_{F})^{-1} \circ\pi^{+}( \hat{D}_{F})^{-1}]
=\int_{M}\int_{|\xi|=1}\texttt{trace}_{\wedge^{\ast}(T^{\ast}M)\otimes F}
  [\sigma_{-n}((\hat{D}^{*}_{F})^{-1} \circ ( \hat{D}_{F})^{-1})]\sigma(\xi)\texttt{d}x+\int_{\partial M} \widetilde{\Psi},
\end{equation}
where
 \begin{eqnarray}
 \widetilde{\Psi}&=&\int_{|\xi'|=1}\int_{-\infty}^{+\infty}\sum_{j,k=0}^{\infty}\sum \frac{(-i)^{|\alpha|+j+k+\ell}}{\alpha!(j+k+1)!}
\texttt{trace}_{\wedge^{\ast}(T^{\ast}M)\otimes F}\Big[\partial_{x_{n}}^{j}\partial_{\xi'}^{\alpha}\partial_{\xi_{n}}^{k}
\sigma_{r}^{+}((\hat{D}^{*}_{F})^{-1})(x',0,\xi',\xi_{n})\nonumber\\
&&\times\partial_{x_{n}}^{\alpha}\partial_{\xi_{n}}^{j+1}\partial_{x_{n}}^{k}\sigma_{l}(( \hat{D}_{F})^{-1})(x',0,\xi',\xi_{n})\Big]
\texttt{d}\xi_{n}\sigma(\xi')\texttt{d}x' ,
\end{eqnarray}
and the sum is taken over $r-k+|\alpha|+\ell-j-1=-n,r\leq-1,\ell\leq-1$.

Locally we can use Theorem 4.4 to compute the interior term of (5.1), then
\begin{eqnarray}
&&\int_{M}\int_{|\xi|=1}{\bf{trace}}_{\wedge^{\ast}(T^{\ast}M)\otimes F}
  [\sigma_{-4}((\hat{D}^{*}_{F})^{-1} \circ ( \hat{D}_{F})^{-1})]\sigma(\xi)\texttt{d}x\nonumber\\
&=&4\pi^{2}\int_{M}{\bf{Tr}}
 \Big[-\frac{s}{12} +\frac{1}{4}[\hat{c}(\omega^{*})-\hat{c}(\omega)]^{2}
         -\frac{1}{4}\hat{c}(\omega^{*})\hat{c}(\omega)-\frac{1}{4}\sum_{j}\nabla_{e_{j}}^{F}\big(\hat{c}(\omega^{*})\big)c(e_{j})\nonumber\\
       &&  +\frac{1}{4}\sum_{j}c(e_{j})\nabla_{e_{j}}^{F}\big(\hat{c}(\omega)\big) \Big]\texttt{d}vol_{M}.
\end{eqnarray}

So we only need to compute $\int_{\partial M} \widetilde{\Psi}$.  In the local coordinates $\{x_{i}; 1\leq i\leq n\}$ and the fixed orthonormal frame
$\{\widetilde{e_{1}},\cdots, \widetilde{e_{n}}\}$, the connection matrix $(\omega_{s,t})$ is defined by
\begin{equation}
\widetilde{\nabla}(\widetilde{e_{1}},\cdots, \widetilde{e_{n}})=(\widetilde{e_{1}},\cdots, \widetilde{e_{n}})(\omega_{s,t}).
\end{equation}

 Let $M$ be a $4$-dimensional compact oriented Riemannian manifold with boundary $\partial M$ and the metric of (6.1).
$ \hat{D}_{F} =d^{F}+\delta^{F}:~C^{\infty}(M,\wedge^{\ast}(T^{\ast}M)\otimes F)\rightarrow C^{\infty}(M,\wedge^{\ast}(T^{\ast}M)\otimes F)$
is the twisted  signature operator. Take the coordinates and
the orthonormal frame as in Section 3.
 Let $\epsilon (\widetilde{e_j*} ),~\iota (\widetilde{e_j*} )$ be the exterior and interior multiplications respectively. Write
 \begin{equation}
c(\widetilde{e_j})=\epsilon (\widetilde{e_j*} )-\iota (\widetilde{e_j*} );~~
\hat{c}(\widetilde{e_j})=\epsilon (\widetilde{e_j*} )+\iota (\widetilde{e_j*} ).
\end{equation}
  We'll compute ${\rm tr}_{\wedge^*(T^*M)\otimes F}$ in the frame $\{e^{\ast}_{i_1}\wedge\cdots\wedge
e^{\ast}_{i_k}|~1\leq i_1<\cdots<i_k\leq 4\}.$ By (3.2) in \cite{Wa3}, we have
\begin{eqnarray}
 \hat{D}_{F}&=&d^{F}+\delta^{F}=\sum_{i=1}^{n}c(e_{i})\nabla^{e}_{e_{i}}-\frac{1}{2}\sum_{i=1}^{n}\hat{c}(e_{i})\omega(F,g^{F})(e_{i})\nonumber\\
    &=&\sum_{i=1}^{n}c(e_{i})\Big(\nabla_{e_{i}}^{\wedge^{\ast}(T^{\ast}M)}\otimes id_{F}+id_{\wedge^{\ast}(T^{\ast}M)} \otimes \nabla^{F,e}_{e_{i}} \Big)
    -\frac{1}{2}\sum_{i=1}^{n}\hat{c}(e_{i})\omega(F,g^{F})(e_{i})\nonumber\\
    &=&\sum^n_{i=1}c(\widetilde{e_i})\Big[\widetilde{e_i}+\frac{1}{4}\sum_{s,t}\omega_{s,t}
(\widetilde{e_i})[\hat{c}(\widetilde{e_s})\hat{c}(\widetilde{e_t})-c(\widetilde{e_s})c(\widetilde{e_t})]\otimes id_{F}
  \nonumber\\
   &&+id_{\wedge^{\ast}(T^{\ast}M)}\otimes \sigma^{F,e}_{i}\Big]-\frac{1}{2}\sum_{i=1}^{n}\hat{c}(e_{i})\omega(F,g^{F})(e_{i}),\\
\hat{D}^{*}_{F}&=&\sum^n_{i=1}c(\widetilde{e_i})\Big[\widetilde{e_i}+\frac{1}{4}\sum_{s,t}\omega_{s,t}
(\widetilde{e_i})[\hat{c}(\widetilde{e_s})\hat{c}(\widetilde{e_t})-c(\widetilde{e_s})c(\widetilde{e_t})]\otimes id_{F}
  \nonumber\\
   &&+id_{\wedge^{\ast}(T^{\ast}M)}\otimes \sigma^{F,e}_{i}\Big]-\frac{1}{2}\sum_{i=1}^{n}\hat{c}(e_{i})\omega^{*}(F,g^{F})(e_{i}).
\end{eqnarray}
Then we obtain
\begin{eqnarray}
\sigma_1(\hat{D}_{F})&=&\sigma_1(\hat{D}^{*}_{F})=\sqrt{-1}c(\xi);\\
\sigma_0(\hat{D}_{F})&=&\sum^n_{i=1}c(\widetilde{e_i})\Big[\frac{1}{4}\sum_{s,t}\omega_{s,t}
(\widetilde{e_i})[\hat{c}(\widetilde{e_s})\hat{c}(\widetilde{e_t})-c(\widetilde{e_s})c(\widetilde{e_t})]\otimes \texttt{id}_{F}
  +id_{\wedge^{\ast}(T^{\ast}M)}\otimes \sigma^{F,e}_{i}\Big]\nonumber\\
   &&-\frac{1}{2}\sum_{i=1}^{n}\hat{c}(e_{i})\omega(F,g^{F})(e_{i});\\
\sigma_0(\hat{D}^{*}_{F})&=&\sum^n_{i=1}c(\widetilde{e_i})\Big[\frac{1}{4}\sum_{s,t}\omega_{s,t}
(\widetilde{e_i})[\hat{c}(\widetilde{e_s})\hat{c}(\widetilde{e_t})-c(\widetilde{e_s})c(\widetilde{e_t})]\otimes \texttt{id}_{F}
  +id_{\wedge^{\ast}(T^{\ast}M)}\otimes \sigma^{F,e}_{i}\Big]\nonumber\\
   &&-\frac{1}{2}\sum_{i=1}^{n}\hat{c}(e_{i})\omega^{*}(F,g^{F})(e_{i}).
\end{eqnarray}

By the composition formula of pseudodifferential operators in Section 2.2.1 of \cite{Wa3}, we have
 \begin{lem}\label{le:31}
The symbol of the  twisted signature operators  $\hat{D}^{*}_{F}, \hat{D}_{F}$ as follows:
\begin{eqnarray}
\sigma_{-1}((\hat{D}_{F})^{-1})&=&\sigma_{-1}((\hat{D}^{*}_{F})^{-1})=\frac{\sqrt{-1}c(\xi)}{|\xi|^{2}}; \\
\sigma_{-2}((\hat{D}_{F})^{-1})&=&\frac{c(\xi)\sigma_{0}(\hat{D}_{F})c(\xi)}{|\xi|^{4}}+\frac{c(\xi)}{|\xi|^{6}}\sum_{j}c(\texttt{d}x_{j})
\Big[\partial_{x_{j}}(c(\xi))|\xi|^{2}-c(\xi)\partial_{x_{j}}(|\xi|^{2})\Big];\\
\sigma_{-2}((\hat{D}^{*}_{F})^{-1})&=&\frac{c(\xi)\sigma_{0}(\hat{D}^{*}_{F})c(\xi)}{|\xi|^{4}}+\frac{c(\xi)}{|\xi|^{6}}\sum_{j}c(\texttt{d}x_{j})
\Big[\partial_{x_{j}}(c(\xi))|\xi|^{2}-c(\xi)\partial_{x_{j}}(|\xi|^{2})\Big].
\end{eqnarray}
\end{lem}

Since $ \widetilde{\Psi}$ is a global form on $\partial M$, so for any fixed point $x_{0}\in\partial M$, we can choose the normal coordinates
$U$ of $x_{0}$ in $\partial M$(not in $M$) and compute $ \widetilde{\Psi}(x_{0})$ in the coordinates $\widetilde{U}=U\times [0,1)$ and the metric
$\frac{1}{h(x_{n})}g^{\partial M}+\texttt{d}x _{n}^{2}$. The dual metric of $g^{\partial M}$ on $\widetilde{U}$ is
$\frac{1}{\tilde{h}(x_{n})}g^{\partial M}+\texttt{d}x _{n}^{2}.$ Write
$g_{ij}^{M}=g^{M}(\frac{\partial}{\partial x_{i}},\frac{\partial}{\partial x_{j}})$;
$g^{ij}_{M}=g^{M}(d x_{i},dx_{j})$, then

\begin{equation}
[g_{i,j}^{M}]=
\begin{bmatrix}\frac{1}{h( x_{n})}[g_{i,j}^{\partial M}]&0\\0&1\end{bmatrix};\quad
[g^{i,j}_{M}]=\begin{bmatrix} h( x_{n})[g^{i,j}_{\partial M}]&0\\0&1\end{bmatrix},
\end{equation}
and
\begin{equation}
\partial_{x_{s}} g_{ij}^{\partial M}(x_{0})=0,\quad 1\leq i,j\leq n-1;\quad g_{i,j}^{M}(x_{0})=\delta_{ij}.
\end{equation}

Let $\{e_{1},\cdots, e_{n-1}\}$ be an orthonormal frame field in $U$ about $g^{\partial M}$ which is parallel along geodesics and
$e_{i}=\frac{\partial}{\partial x_{i}}(x_{0})$, then $\{\widetilde{e_{1}}=\sqrt{h(x_{n})}e_{1}, \cdots,
\widetilde{e_{n-1}}=\sqrt{h(x_{n})}e_{n-1},\widetilde{e_{n}}=\texttt{d}x_{n}\}$ is the orthonormal frame field in $\widetilde{U}$ about $g^{M}.$
Locally $\wedge^{\ast}(T^{\ast}M)|\widetilde{U}\cong \widetilde{U}\times\wedge^{*}_{C}(\frac{n}{2}).$ Let $\{f_{1},\cdots,f_{n}\}$ be the orthonormal basis of
$\wedge^{*}_{C}(\frac{n}{2})$. Take a spin frame field $\sigma: \widetilde{U}\rightarrow Spin(M)$ such that
$\pi\sigma=\{\widetilde{e_{1}},\cdots, \widetilde{e_{n}}\}$ where $\pi: Spin(M)\rightarrow O(M)$ is a double covering, then
$\{[\sigma, f_{i}], 1\leq i\leq 4\}$ is an orthonormal frame of $\wedge^{\ast}(T^{\ast}M)|_{\widetilde{U}}.$ In the following,
since the global form $ \widetilde{\Psi}$
is independent of the choice of the local frame, so we can compute $\texttt{tr}_{\wedge^{\ast}(T^{\ast}M)}$ in the frame
$\{[\sigma, f_{i}], 1\leq i\leq 4\}$.
Let $\{E_{1},\cdots,E_{n}\}$ be the canonical basis of $R^{n}$ and
$c(E_{i})\in cl_{C}(n)\cong \texttt{Hom}(\wedge^{*}_{C}(\frac{n}{2}),\wedge^{*}_{C}(\frac{n}{2}))$ be the Clifford action. By \cite{Wa3}, then

\begin{equation}
c(\widetilde{e_{i}})=[(\sigma,c(E_{i}))]; \quad c(\widetilde{e_{i}})[(\sigma, f_{i})]=[\sigma,(c(E_{i}))f_{i}]; \quad
\frac{\partial}{\partial x_{i}}=[(\sigma,\frac{\partial}{\partial x_{i}})],
\end{equation}
then we have $\frac{\partial}{\partial x_{i}}c(\widetilde{e_{i}})=0$ in the above frame. By Lemma 2.2 in \cite{Wa3}, we have

\begin{lem}\label{le:32}
With the metric $g^{M}$ on $M$ near the boundary
\begin{eqnarray}
\partial_{x_j}(|\xi|_{g^M}^2)(x_0)&=&\left\{
       \begin{array}{c}
        0,  ~~~~~~~~~~ ~~~~~~~~~~ ~~~~~~~~~~~~~{\rm if }~j<n; \\[2pt]
       h'(0)|\xi'|^{2}_{g^{\partial M}},~~~~~~~~~~~~~~~~~~~~~{\rm if }~j=n.
       \end{array}
    \right. \\
\partial_{x_j}[c(\xi)](x_0)&=&\left\{
       \begin{array}{c}
      0,  ~~~~~~~~~~ ~~~~~~~~~~ ~~~~~~~~~~~~~{\rm if }~j<n;\\[2pt]
\partial x_{n}(c(\xi'))(x_{0}), ~~~~~~~~~~~~~~~~~{\rm if }~j=n,
       \end{array}
    \right.
\end{eqnarray}
where $\xi=\xi'+\xi_{n}\texttt{d}x_{n}$
\end{lem}
Then an application of Lemma 2.3 in \cite{Wa3} shows
\begin{lem}
The symbol of the  twisted signature operators  $\hat{D}^{*}_{F},   \hat{D}_{F}$
\begin{eqnarray}
\sigma_{0}(\hat{D}^{*}_{F})&=&-\frac{3}{4}h'(0)c(dx_n)
+\frac{1}{4}h'(0)\sum^{n-1}_{i=1}c(\widetilde{e_i})\hat{c}(\widetilde{e_n})\hat{c}(\widetilde{e_i})(x_0)\otimes id_{F}  \nonumber\\
  &&+\sum^n_{i=1}c(\widetilde{e_i})\sigma^{F,e}_{i}
   -\frac{1}{2}\sum_{i=1}^{n}\hat{c}(e_{i})\omega^{*}(F,g^{F})(e_{i});\\
 \sigma_{0}(\hat{D}_{F})&=&-\frac{3}{4}h'(0)c(dx_n)
+\frac{1}{4}h'(0)\sum^{n-1}_{i=1}c(\widetilde{e_i})\hat{c}(\widetilde{e_n})\hat{c}(\widetilde{e_i})(x_0)\otimes id_{F}  \nonumber\\
  &&+\sum^n_{i=1}c(\widetilde{e_i})\sigma^{F,e}_{i}
   -\frac{1}{2}\sum_{i=1}^{n}\hat{c}(e_{i})\omega(F,g^{F})(e_{i})
\end{eqnarray}
\end{lem}

Now we can compute $ \widetilde{\Psi}$ (see formula (5.2) for definition of $ \widetilde{\Psi}$), since the sum is taken over $-r-\ell+k+j+|\alpha|=3,
 \ r, \ell\leq-1$, then we have the following five cases:

\textbf{Case a(I)}: \ $r=-1, \ \ell=-1, \ k=j=0, \ |\alpha|=1$

From (5.2), we have
\begin{equation}
\text{ Case \ a(\text{I})}=-\int_{|\xi'|=1}\int_{-\infty}^{+\infty}\sum_{|\alpha|=1}\text{trace}
[\partial_{\xi'}^{\alpha}\pi_{\xi_{n}}^{+}\sigma_{-1}((\hat{D}^{*}_{F} )^{-1})
\partial_{x'}^{\alpha}\partial_{\xi_{n}}\sigma_{-1}((\hat{D}_{F})^{-1})](x_{0})\texttt{d}\xi_{n}\sigma(\xi')\texttt{d}x' .
\end{equation}
By Lemma 5.2, for $j<n$
\begin{equation}
\partial_{x_j}\sigma_{-1}(( \hat{D}_{F})^{-1})(x_0)=\partial_{x_j}\left(\frac{\sqrt{-1}c(\xi)}{|\xi|^2}\right)(x_0)=
\frac{\sqrt{-1}\partial_{x_i}[c(\xi)](x_0)}{|\xi|^2}
-\frac{\sqrt{-1}c(\xi)\partial_{x_i}(|\xi|^2)(x_0)}{|\xi|^4}=0,
\end{equation}
so Case a(I) vanishes.

\textbf{Case a(II)}: \ $r=-1, \ \ell=-1, \ k=|\alpha|=0, \ j=1$

From (5.2), we have
\begin{equation}
\text{ Case \ a(\text{II})}=-\frac{1}{2}\int_{|\xi'|=1}\int_{-\infty}^{+\infty}\text{trace}[\partial_{x_{n}}\pi_{\xi_{n}}^{+}
\sigma_{-1}((\hat{D}^{*}_{F})^{-1})\partial_{\xi_{n}}^{2}\sigma_{-1}(( \hat{D}_{F})^{-1})](x_{0})\texttt{d}\xi_{n}\sigma(\xi')\texttt{d}x'.
\end{equation}
Similar to (2.2.18) in \cite{Wa3}, we have
\begin{equation}
\partial_{x_{n}}\pi_{\xi_{n}}^{+}\sigma_{-1}((\hat{D}^{*}_{F})^{-1})(x_{0})|_{|\xi'|=1}=\frac{\partial_{x_{n}}[c(\xi')](x_{0})}{2(\xi_{n}-i)}
+\sqrt{-1}h'(0)[\frac{ic(\xi')}{4(\xi_{n}-i)}+\frac{c(\xi')+ic(\texttt{d}x_{n})}{4(\xi_{n}-i)^{2}}];
\end{equation}
and
\begin{equation}
\partial_{\xi_{n}}^{2}\sigma_{-1}((\hat{D}_{F})^{-1})=\sqrt{-1}(-\frac{6\xi_{n}c(\texttt{d}x_{n})+2c(\xi')}{|\xi|^{4}}
+\frac{8\xi_{n}^{2}c(\xi)}{|\xi|^{6}}).
\end{equation}

 Let $dim F=l$, by the relation of the Clifford action and $\texttt{tr}AB=\texttt{tr}BA$,  then
\begin{eqnarray}
&&\texttt{tr}[c(\xi')c(\texttt{d}x_{n})]=0; \ \texttt{tr}[c(\texttt{d}x_{n})^{2}]=-16l;\
\texttt{tr}[c(\xi')^{2}](x_{0})|_{|\xi'|=1}=-16l;\nonumber\\
&&\texttt{tr}[\partial_{x_{n}}[c(\xi')]c(\texttt{d}x_{n})]=0; \
\texttt{tr}[\partial_{x_{n}}c(\xi')\times c(\xi')](x_{0})|_{|\xi'|=1}=-8lh'(0).
\end{eqnarray}
For more trace expansions, we can see \cite{GS}.

Then
\begin{equation}
\text{trace}\Big[\partial_{x_{n}}\pi_{\xi_{n}}^{+}
\sigma_{-1}((\hat{D}^{*}_{F})^{-1})\partial_{\xi_{n}}^{2}\sigma_{-1}(( \hat{D}_{F})^{-1})\Big](x_{0})
=\frac{8lih'(0)}{(\xi_{n}-i)^{2}(\xi_{n}+i)^{3}}.
\end{equation}
Therefore
\begin{eqnarray}
{\rm case~ a(II)}&=&-\int_{|\xi'|=1}\int^{+\infty}_{-\infty}\frac{4lih'(0)(\xi_n-i)^2}
{(\xi_n-i)^4(\xi_n+i)^3}d\xi_n\sigma(\xi')dx'\nonumber\\
&=&-4lih'(0)\Omega_3\int_{\Gamma^+}\frac{1}{(\xi_n-i)^2(\xi_n+i)^3}d\xi_ndx'\nonumber\\
&=&-4lih'(0)\Omega_32\pi i\Big[\frac{1}{(\xi_n+i)^3}\Big]^{(1)}|_{\xi_n=i}dx'\nonumber\\
&=&-\frac{3l}{2}\pi h'(0)\Omega_3dx',
\end{eqnarray}
where $\Omega_{3}$ is the canonical volume of $S^{3}.$

\textbf{Case a(III)}: \ $r=-1, \ \ell=-1, \ j=|\alpha|=0, \ k=1$

From (5.2), we have
\begin{equation}
\text{ Case \ a(\text{III}) }=-\frac{1}{2}\int_{|\xi'|=1}\int_{-\infty}^{+\infty}\text{trace}[\partial_{\xi_{n}}\pi_{\xi_{n}}^{+}
\sigma_{-1}((\hat{D}^{*}_{F})^{-1})\partial_{\xi_{n}}\partial_{x_{n}}\sigma_{-1}((\hat{D}_{F})^{-1})](x_{0})\texttt{d}\xi_{n}\sigma(\xi')\texttt{d}x' .
 \end{equation}
Similar to (2.2.27) in \cite{Wa3}, we have
\begin{equation}
\partial_{\xi_{n}}\pi_{\xi_{n}}^{+}\sigma_{-1}((\hat{D}^{*}_{F})^{-1})(x_{0})|_{|\xi'|=1}=-\frac{c(\xi')+ic(\texttt{d}x_{n})}{2(\xi_{n}-i)^{2}},
\end{equation}
and
\begin{equation}
\partial_{\xi_{n}}\partial_{x_{n}}\sigma_{-1}((\hat{D}_{F})^{-1})(x_{0})|_{|\xi'|=1}
=-\sqrt{-1}h'(0)\Big[\frac{c(\texttt{d}x_{n})}{|\xi|^{4}}-4\xi_{n}\frac{c(\xi')+\xi_{n}c(\texttt{d}x_{n})}{|\xi|^{6}}\Big]
-\frac{2\sqrt{-1}\xi_{n}\partial_{x_{n}}c(\xi')(x_{0})}{|\xi|^{4}}.
\end{equation}
Combining (5.30) and (5.31), we obtain
\begin{equation}
\text{trace}\Big[\partial_{\xi_{n}}\pi_{\xi_{n}}^{+}\sigma_{-1}((\hat{D}^{*}_{F})^{-1})
\partial_{\xi_{n}}\partial_{x_{n}}\sigma_{-1}((\hat{D}_{F})^{-1})\Big](x_{0})
=\frac{8lh'(0)(i-2\xi_{n}-i\xi_{n}^{2})}{(\xi_{n}-i)^{4}(\xi_{n}+i)^{3}}.
\end{equation}
Then
\begin{equation}
\text{ Case \ a(\text{III}) }=\frac{3l}{2}\pi h'(0)\Omega_{3}\texttt{d}x',
\end{equation}
where $\Omega_{3}$ is the canonical volume of $S^{3}.$
Thus the sum of Case a(\texttt{II}) and Case a(\texttt{III}) is zero.

\textbf{Case b}: \ $r=-2, \ \ell=-1, \ k=j=|\alpha|=0$

By (5.2), we get
\begin{equation}
\text{ Case \ b}=-i\int_{|\xi'|=1}\int_{-\infty}^{+\infty}\text{trace}[\pi_{\xi_{n}}^{+}\sigma_{-2}((\hat{D}^{*}_{F})^{-1})
                 \partial_{\xi_{n}}\sigma_{-1}((\hat{D}_{F})^{-1})](x_{0})\texttt{d}\xi_{n}\sigma(\xi')\texttt{d}x'.
\end{equation}
Then an application of Lemma 5.2 shows
\begin{eqnarray}
\sigma_{-2}((\hat{D}^{*}_{F})^{-1})(x_{0})&=&\frac{c(\xi)\sigma_{0}(D^{F})(x_{0})c(\xi)}{|\xi|^{4}}+\frac{c(\xi)}{|\xi|^{6}}\sum_{j}c(dx_{j})
                    \Big[\partial_{x_{j}}(c(\xi))|\xi|^{2}-c(\xi)\partial_{x_{j}}(|\xi|^{2})\Big](x_{0})\nonumber\\
                   &=&\frac{c(\xi)\sigma_{0}(D^{F})(x_{0})c(\xi)}{|\xi|^{4}}+
                  + \frac{c(\xi)}{|\xi|^{6}}c(dx_{n})\Big[\partial_{ x_{n}}(c(\xi'))(x_{0})-c(\xi)h'(0)|\xi'|^{2}_{g^{\partial M}}\Big].
\end{eqnarray}
Hence,
\begin{equation}
\pi_{\xi_{n}}^{+}\sigma_{-2}((\hat{D}^{*}_{F})^{-1})(x_{0}):=A_{1}+A_{2},
\end{equation}
where
\begin{eqnarray}
A_{1}&=&
-\frac{h'(0)}{2}\left[\frac{c(dx_n)}{4i(\xi_n-i)}+\frac{c(dx_n)-ic(\xi')}{8(\xi_n-i)^2}
+\frac{3\xi_n-7i}{8(\xi_n-i)^3}[ic(\xi')-c(dx_n)]\right]  \nonumber\\
&&+\frac{-1}{4(\xi_n-i)^2}\Big[(2+i\xi_n)c(\xi')\alpha_0c(\xi')+i\xi_nc(dx_n)\alpha_0c(dx_n)+(2+i\xi_n)c(\xi')c(dx_n)\partial_{x_n}c(\xi')\nonumber\\
&&~~~~+ic(dx_n)\alpha_0c(\xi')+ic(\xi')\alpha_0c(dx_n)-i\partial_{x_n}c(\xi')\Big];\\
A_{2}&=&\frac{-1}{4(\xi_n-i)^2}\Big[(2+i\xi_n)c(\xi')\beta_0c(\xi')+i\xi_nc(dx_n)\beta_0c(dx_n)+ic(dx_n)\beta_0c(\xi')\nonumber\\
 &&+ic(\xi')\beta_0c(dx_n)\Big]
\end{eqnarray}
and
\begin{eqnarray}
\alpha_0&=&-\frac{3}{4}h'(0)c(dx_n)
    +\frac{1}{4}h'(0)\sum^{n-1}_{i=1}c(\widetilde{e_i})\hat{c}(\widetilde{e_n})\hat{c}(\widetilde{e_i})(x_0)\otimes id_{F}
    :=-\frac{3}{4}h'(0)c(dx_n)+p_0\otimes id_{F},\\
\beta_0&=&\sum^n_{i=1}c(\widetilde{e_i})\sigma^{F,e}_{i}
   -\frac{1}{2}\sum_{i=1}^{n}\hat{c}(e_{i})\omega^{*}(F,g^{F})(e_{i}).
\end{eqnarray}
On the other hand,
\begin{equation}
\partial_{\xi_{n}}\sigma_{-1}((\hat{D}_{F})^{-1})=\frac{-2 i \xi_{n}c(\xi')}{(1+\xi_{n}^{2})^{2}}+\frac{i(1- \xi_{n}^{2})c(\texttt{d}x_{n})}
{(1+\xi_{n}^{2})^{2}}.
\end{equation}
For the signature operator case,
\begin{equation}
{\rm tr}[c(\xi')\alpha_0c(\xi')c(dx_n)](x_0)={\rm tr}[\alpha_0c(\xi')c(dx_n)c(\xi')](x_0)=|\xi'|^2{\rm tr}[\alpha_0c(dx_n)],
\end{equation}
 and
 \begin{eqnarray}
c(dx_n)p_0(x_0)&=&-\frac{1}{4}h'(0)\sum^{n-1}_{i=1}c(\tilde{e}_i)\hat{c}(\tilde{e}_i)c(\widetilde{e_n})\hat{c}(\widetilde{e_n})\nonumber\\
&=&-\frac{1}{4}h'(0)\sum^{n-1}_{i=1}[\epsilon ({\widetilde{e_i*}} )\iota
({\widetilde{e_i*}} )-\iota(\widetilde{e_i*})\epsilon(\widetilde{e_i*})][\epsilon ({\widetilde{e_n*}} )\iota
({\widetilde{e_n*}} )-\iota(\widetilde{e_n*})\epsilon(\widetilde{e_n*})].
\end{eqnarray}
 By Section 3 in \cite{Wa3}, then
 \begin{eqnarray}
&& {\rm tr}_{\wedge^m(T^*M)} \{[\epsilon ({e_i*} )\iota ({e_i*} )-\iota(e_i*)\epsilon(e_i*)]
[\epsilon ({e_n*} )\iota ({e_n*} )-\iota(e_n*)\epsilon(e_n*)]\}\nonumber\\
&=&a_{n,m}\langle e_i*,e_n* \rangle^{2}+b_{n,m}|e_i*|^2|e_n*|^2=b_{n,m},
 \end{eqnarray}
 where $b_{4,m}=\left(\begin{array}{lcr}
  \ \ 2 \\
    \  m-2
\end{array}\right)+\left(\begin{array}{lcr}
  \ \ 2 \\
    \  m
\end{array}\right)-2\left(\begin{array}{lcr}
  \ \ 2 \\
    \  m-1
\end{array}\right).$
Then
\begin{eqnarray}
{\rm tr}_{\wedge^*(T^*M)} \{[\epsilon ({\widetilde{e_i*}} )\iota ({\widetilde{e_i*}} )-\iota(\widetilde{e_i*})\epsilon(\widetilde{e_i*})]
[\epsilon ({\widetilde{e_n*}} )\iota ({\widetilde{e_n*}} )-\iota(\widetilde{e_n*})\epsilon(\widetilde{e_n*})]\}
=\sum_{m=0}^4b_{4,m}=0.
\end{eqnarray}
Hence in this case,
\begin{eqnarray}
{\rm tr}_{\wedge^*(T^*M)}[c(dx_n)p_0(x_0)]=0.
\end{eqnarray}
We note that $\int_{|\xi'|=1}\xi_1\cdots\xi_{2q+1}\sigma(\xi')=0$, then
${\rm tr}_{\wedge^*(T^*M)}[c(\xi')p_0(x_0)]$ has no contribution for computing Case b.

From (2.2.39), (2.2.41) and (2.2.42) in \cite{Wa3}, we have
\begin{equation}
-i\int_{|\xi'|=1}\int_{-\infty}^{+\infty}\text{trace}[A_{1}\times
                 \partial_{\xi_{n}}\sigma_{-1}((\hat{D}_{F})^{-1})](x_{0})\texttt{d}\xi_{n}\sigma(\xi')\texttt{d}x'
                 =\frac{9l}{2}\pi h'(0)\Omega_{3}\texttt{d}x'.
\end{equation}
Combining (5.38) and (5.41), we obtain
\begin{equation}
\text{trace}[A_{2}\times\partial_{\xi_{n}}\sigma_{-1}((\hat{D}_{F})^{-1})](x_{0})
=\frac{-1-2i\xi_{n}+\xi_{n}^{2}}{2(\xi_{n}+i)^{2}(\xi_{n}-i)^{4}}\texttt{tr}[\beta_0c(\xi')]
 +\frac{-i+2\xi_{n}-i\xi_{n}^{2}}{2(\xi_{n}+i)^{2}(\xi_{n}-i)^{4}}\texttt{tr}[\beta_0c(\texttt{d}x_{n})].
\end{equation}
By the relation of the Clifford action and $\texttt{tr}AB=\texttt{tr}BA$, then we have the equalities
\begin{equation}
\texttt{tr}[c(\widetilde{e_{i}})c(\texttt{d}x_{n})]=0, i<n; \ \texttt{tr}[c(\widetilde{e_{i}})c(\texttt{d}x_{n})]=-16l, i=n;
~~\texttt{tr}[\hat{c}(\widetilde{e_{i}})c(\xi')]=\texttt{tr}[\hat{c}(\widetilde{e_{i}})c(\texttt{d}x_{n})]=0.
\end{equation}
Then $\texttt{tr}[\beta_0c(\xi')]$ has no contribution for computing Case b.

From (5.48) and (5.49), we have
\begin{eqnarray}
&&-i\int_{|\xi'|=1}\int_{-\infty}^{+\infty}\text{trace}[A_{2}\times
                 \partial_{\xi_{n}}\sigma_{-1}((\hat{D}_{F})^{-1})](x_{0})\texttt{d}\xi_{n}\sigma(\xi')\texttt{d}x'\nonumber\\
&=&2i\Omega_{3}\int_{\Gamma^{+}}\frac{-i+2\xi_{n}-i\xi_{n}^{2}}{2(\xi_{n}+i)^{2}(\xi_{n}-i)^{4}}
\texttt{tr}[id_{\wedge^{\ast}(T^{\ast}M)}\otimes \sigma^{F,e}_{n}]\texttt{d}\xi_{n}\texttt{d}x' \nonumber\\
&=&2i \frac{2\pi i }{3!}\Omega_{3}\Big[\frac{-i+2\xi_{n}-i\xi_{n}^{2}}{2(\xi_{n}+i)^{2}}\Big]^{(1)}|_{\xi_{n}=i}
  \texttt{tr}[id_{\wedge^{\ast}(T^{\ast}M)}\otimes \sigma^{F,e}_{n}]\texttt{d}x'  \nonumber\\
&=&-4\pi \texttt{tr}_{F}[\sigma^{F,e}_{n}]\Omega_{3}\texttt{d}x'.
\end{eqnarray}
Combining (5.47) and (5.50), we have
\begin{equation}
\texttt{case}\ b=\Big[\frac{9}{2}lh'(0)-4 \texttt{tr}_{F}[\sigma^{F,e}_{n}]\Big]\pi \Omega_{3}\texttt{d}x'.
\end{equation}

\textbf{Case c}: \ $r=-1, \ \ell=-2, \ k=j=|\alpha|=0$

From (5.2), we have
\begin{equation}
 \text{case \ c}=-i\int_{|\xi'|=1}\int_{-\infty}^{+\infty}\text{trace}[\pi_{\xi_{n}}^{+}\sigma_{-1}((\hat{D}^{*}_{F})^{-1})
 \partial_{\xi_{n}}\sigma_{-2}((\hat{D}_{F})^{-1})](x_{0})\texttt{d}\xi_{n}\sigma(\xi')\texttt{d}x'  .
 \end{equation}
From (5.13) we obtain
\begin{equation}
\pi_{\xi_{n}}^{+}\sigma_{-1}((\hat{D}^{*}_{F})^{-1})=\frac{c(\xi')+ic(\texttt{d}x_{n})}{2(\xi_{n}-i)}.
\end{equation}
Hence,
\begin{equation}
 \partial_{\xi_{n}}\sigma_{-2}((\hat{D}_{F})^{-1})(x_{0}):=B_{1}+B_{2},
\end{equation}
where
\begin{eqnarray}
B_{1}&=&\frac{1}{(1+\xi_n^2)^3}\Big[(2\xi_n-2\xi_n^3)c(dx_n)\alpha_0c(dx_n)+(1-3\xi_n^2)c(dx_n)\alpha_0c(\xi')\nonumber\\
    &&+ (1-3\xi_n^2)c(\xi')\alpha_0c(dx_n)-4\xi_nc(\xi')\alpha_0 c(\xi')+(3\xi_n^2-1)\partial_{x_n}c(\xi')
    -4\xi_nc(\xi')c(dx_n)\partial_{x_n}c(\xi')\nonumber\\
    &&+2h'(0)c(\xi')+2h'(0)\xi_nc(dx_n)\Big]
    +6\xi_nh'(0)\frac{c(\xi)c(dx_n)c(\xi)}{(1+\xi^2_n)^4};\\
B_{2}&=&\frac{1}{(1+\xi_n^2)^3}\Big[(2\xi_n-2\xi_n^3)c(dx_n)\beta_0c(dx_n)+(1-3\xi_n^2)c(dx_n)\beta_0c(\xi')\nonumber\\
    &&+(1-3\xi_n^2)c(\xi')\beta_0c(dx_n)-4\xi_nc(\xi')\beta_0c(\xi')\Big].
\end{eqnarray}
Similar to \textbf{Case b}, we have
\begin{equation}
-i\int_{|\xi'|=1}\int_{-\infty}^{+\infty}\text{trace}[\pi_{\xi_{n}}^{+}\sigma_{-1}((\hat{D}^{*}_{F})^{-1})\times B_{1}]
               (x_{0})\texttt{d}\xi_{n}\sigma(\xi')\texttt{d}x'
                 =-\frac{9l}{2}\pi h'(0)\Omega_{3}\texttt{d}x'.
\end{equation}
From (5.53) and (5.56),  we obtain
\begin{eqnarray}
\text{trace}[\pi_{\xi_{n}}^{+}\sigma_{-1}((\hat{D}^{*}_{F})^{-1})\times B_{2}](x_{0})
&=&\frac{-i+3\xi_{n}+3 i\xi_{n}^{2}-\xi_{n}^{3}}{(\xi_{n}+i)^{3}(\xi_{n}-i)^{4}}\texttt{tr}[\beta_0c(\xi')]\nonumber\\
 &&+\frac{-1-3i\xi_{n}+3\xi_{n}^{2}+i\xi_{n}^{3}}{(\xi_{n}+i)^{3}(\xi_{n}-i)^{4}}\texttt{tr}[\beta_0c(\texttt{d}x_{n})].
\end{eqnarray}
Hence in this case,
\begin{eqnarray}
&&-i\int_{|\xi'|=1}\int_{-\infty}^{+\infty}\text{trace}[\pi_{\xi_{n}}^{+}\sigma_{-1}((\hat{D}^{*}_{F})^{-1})\times B_{2}]
                  (x_{0})\texttt{d}\xi_{n}\sigma(\xi')\texttt{d}x'\nonumber\\
&=&2i\Omega_{3}\int_{\Gamma^{+}}\frac{-1-3i\xi_{n}+3\xi_{n}^{2}+i\xi_{n}^{3}}{(\xi_{n}+i)^{3}(\xi_{n}-i)^{4}}
 \texttt{tr}[id_{\wedge^{\ast}(T^{\ast}M)}\otimes \sigma^{F,e}_{n}]\texttt{d}\xi_{n}\texttt{d}x' \nonumber\\
&=&2i \frac{2\pi i}{3!} \Omega_{3}\Big[\frac{-1-3i\xi_{n}+3\xi_{n}^{2}+i\xi_{n}^{3}}{(\xi_{n}+i)^{3}}\Big]^{(1)}|_{\xi_{n}=i}
  \texttt{tr}[id_{\wedge^{\ast}(T^{\ast}M)}\otimes \sigma^{F,e}_{n}]\texttt{d}x'  \nonumber\\
&=&4\pi \texttt{tr}_{F}[\sigma^{F,e}_{n}]\Omega_{3}\texttt{d}x'.
\end{eqnarray}
Combining (5.57) and (5.59), we obtain
\begin{equation}
\texttt{case}\ c=\Big[-\frac{9}{2}lh'(0)+4\texttt{tr}_{F}[\sigma^{F,e}_{n}]\Big]\pi \Omega_{3}\texttt{d}x'.
\end{equation}
Now $ \widetilde{\Psi}$  is the sum of the \textbf{case (a, b, c)}, so
\begin{equation}
\sum \textbf{case a, b , c}=0.
\end{equation}
Hence we conclude that

\begin{thm}  Let  $M$ be a $4$-dimensional compact oriented Riemaniann manifold
and the metric $g^M$ as above, let $\hat{D}^{*}_{F}, \hat{D}_{F}$ be the twisted signature operators on
 $C^{\infty}(M,\wedge^{\ast}(T^{\ast}M)\otimes F)$,
 then
\begin{eqnarray}
\widetilde{Wres}[\pi^+(\hat{D}^{*}_{F})^{-1}\pi^+(\hat{D}_{F})^{-1}]&=&4\pi^{2}\int_{M}{\bf{Tr}}
 \Big[-\frac{s}{12} +\frac{1}{4}[\hat{c}(\omega^{*})-\hat{c}(\omega)]^{2}
         -\frac{1}{4}\hat{c}(\omega^{*})\hat{c}(\omega)\nonumber\\
       && -\frac{1}{4}\sum_{j}\nabla_{e_{j}}^{F}\big(\hat{c}(\omega^{*})\big)c(e_{j})
        +\frac{1}{4}\sum_{j}c(e_{j})\nabla_{e_{j}}^{F}\big(\hat{c}(\omega)\big) \Big]\texttt{d}vol_{M}.
\end{eqnarray}
where $s$ be the scalar curvature of $M$.
\end{thm}

Now we give a Kastler-Kalau-Walze theorem for  $\hat{D}^{2}_{F}$.
let $\Delta^{e}$ be the Bochner laplacian
 \begin{equation}
\Delta^{e}=\sum_{i=1}^{n}\Big[(\nabla^{e}_{e_{i}})^{2}-\nabla^{e}_{ \nabla^{TM}_{e_{i}}e_{i}} \Big].
\end{equation}
 From Theorem 1.1 in \cite{ZW}, we can state
the following Lichnerowicz type formula for $D^{2}_{F}$.

 \begin{thm}\cite{ZW}
The following identity holds
\begin{eqnarray}
\hat{D}^{2}_{F}&=& -\Delta^{e}+\frac{s}{4}-\frac{1}{8}\sum_{i=1}^{n}c(e_{i})c(e_{j})(\omega(F,g^{F}))^{2}(e_{i},e_{j}) \nonumber\\
&&+\frac{1}{8}\sum_{i,j=1}^{n}\langle R^{TM}(e_{i},e_{j})e_{i},e_{j} \rangle c(e_{i})c(e_{j})\hat{c}(e_{i})\hat{c}(e_{j})\nonumber\\
&&+\frac{1}{4}\sum_{j=1}^{n}(\omega(F,g^{F})(e_{j}))^{2}+\frac{1}{8}\sum_{i,j=1}^{n}\hat{c}(e_{i})\hat{c}(e_{j})
 (\omega(F,g^{F}))^{2}(e_{i},e_{j})\nonumber\\
&&-\frac{1}{4}\sum_{i,j=1}^{n}c(e_{i})\hat{c}(e_{j})\Big(\nabla^{F}_{e_{i}}\omega(F,g^{F})(e_{j})
+\nabla^{F}_{e_{j}}\omega(F,g^{F})(e_{i})   \Big)\nonumber\\
&:=&-\Delta^{e}+E.
\end{eqnarray}
\end{thm}
Therefore
 \begin{equation}
\texttt{tr}[E]=\texttt{tr}\Big[\frac{s}{4} +\frac{1}{2}\big(\omega(F,g^{F})(e_{i})\big)^{2} \Big].
\end{equation}
 Then we obtain

\begin{thm}  Let  $M$ be a $4$-dimensional compact oriented Riemaniann manifold without boundary,
and $\hat{D}_{F}=d^{F}+\delta^{F}$ be the twisted signature operator on  $C^{\infty}(M,\wedge^{\ast}(T^{\ast}M)\otimes F)$,
 then
 \begin{equation}
Wres[\hat{D}^{-2}_{F}]=2\pi^{2}\int_M{\bf{tr}}\Big[\frac{5}{6}s +\big(\omega(F,g^{F})(e_{i})\big)^{2} \Big]{\rm dvol}_M.
\end{equation}
\end{thm}

\begin{thm}  Let  $M$ be a $4$-dimensional compact oriented Riemaniann manifold with boundary,
and $\hat{D}_{F}=d^{F}+\delta^{F}$ be the twisted signature operator on  $C^{\infty}(M,\wedge^{\ast}(T^{\ast}M)\otimes F)$,
 then
 \begin{equation}
Wres[(\pi^+(\hat{D}_{F})^{-1})^2]=2\pi^{2}\int_M{\bf{tr}}\Big[\frac{5}{6}s +\big(\omega(F,g^{F})(e_{i})\big)^{2} \Big]{\rm dvol}_M.
\end{equation}
\end{thm}

\section*{ Acknowledgements}
This work was supported by Fok Ying Tong Education Foundation under Grant No. 121003,  NSFC. 11271062 and NCET-13-0721.  The authors
also thank the referee for his (or her) careful reading and helpful comments.


\section*{References}
\begin{thebibliography}{00}
\bibitem{Gu} V. W. Guillemin.: A new proof of Weyl's formula on the asymptotic distribution of eigenvalues. Adv. Math. 55, no. 2, 131-160, (1985).
\bibitem{Wo} M.Wodzicki.: local invariants of spectral asymmetry. Invent. Math. 75(1), 143-178, (1995).
\bibitem{MA} M. Adler.: On a trace functional for formal pseudo-differential operators and the symplectic
structure of Korteweg-de Vries type equations, Invent. Math. 50, 219-248,(1979).
\bibitem{Co1} A. Connes.: Quantized calculus and applications.  XIth International Congress of Mathematical Physics(Paris,1994),
 Internat Press, Cambridge, MA, 15-36, (1995).
\bibitem{Co2} A. Connes.: The action functinal in Noncommutative geometry. Comm. Math. Phys. 117, 673-683, (1998).
\bibitem{Ka} D. Kastler.: The Dirac Operator and Gravitation. Comm. Math. Phys. 166, 633-643, (1995).
\bibitem{KW} W. Kalau and M. Walze.: Gravity, Noncommutative geometry and the Wodzicki residue. J. Geom. Phys.  16, 327-344, (1995).
\bibitem{Ac} T. Ackermann.: A note on the Wodzicki residue. J. Geom. Phys. 20, 404-406, (1996).
\bibitem{FGLS} B. V. Fedosov, F. Golse, E. Leichtnam, E. Schrohe.: The noncommutative residue for manifolds with boundary.
J. Funct. Anal. 142, 1-31, (1996).
\bibitem{Wa3} Y. Wang.: Gravity and the Noncommutative Residue for Manifolds with Boundary. Lett. Math. Phy. 80, 37-56, (2007).
\bibitem{Wa4} Y. Wang.: Lower-Dimensional Volumes and Kastler-kalau-Walze Type Theorem for Manifolds with Boundary .
      Commun. Theor. Phys. Vol 54, 38-42, (2010).
\bibitem{WW} J. Wang and  Y. Wang.:  Nonminimal operators and non-commutative residue, J. Math. Phys. 53, 072503 (2012).
\bibitem{WW1} J. Wang and  Y. Wang.: Noncommutative residue and sub-Dirac operators for foliations, J. Math. Phys.  54, 012501 (2013).
 \bibitem{BZ} J. M. Bismut and W. Zhang.: An Extension of a theorem by Cheeger and M$\ddot{u}$ller, Ast$\acute{e}$risque, No. 205,
 paris, (1992).
\bibitem{MZ} X.N. MA and W. ZHANG.:  $\eta$-Invariant and Flat Vector Bundles. Chin. Ann. Math. 27B(1), 67-72, (2006).
\bibitem{ZW} W. Zhang.: Sub-signature operators, $ \eta$ invariants and a Riemann-Roch theorem for flat vector bundles.
 Chin. Ann. Math. 25B: 1, 7-36, (2004).
 \bibitem{PBG} P. B. Gilkey, Invariance Theory, the Heat Equation, and the Atiyah-Singer Index Theorem, vol. 11 of Mathematics Lecture
Series, 1984.
\bibitem{AT} T. Ackermann and J. Tolksdorf.:  A generalized Lichnerowicz formula, the Wodzicki residue and gravity.
J. Geom. Phys. 19, 143-150,(1996).
\bibitem{Wa1} Y. Wang.: Diffential forms and the Wodzicki residue for Manifolds with Boundary. J. Geom. Phys.  56, 731-753, (2006).
\bibitem{GS} G. Grubb, E. Schrohe.: Trace expansions and the noncommutative residue for manifolds with boundary.
J. Reine Angew. Math. 536, 167-207, (2001).
\end{thebibliography}
\end{document}